\documentclass[aps,prb,superscriptaddress,letterpaper,amsmath,amssymb,twocolumn,10pt,floatfix]{revtex4-2}

\usepackage{graphicx}
\usepackage[ansinew]{inputenc} % set .tex file encoding to ANSI
\usepackage{array} % used for improved array and tabular environments
\usepackage{xcolor} % modern handling of color in LaTeX
\usepackage{amsxtra}
\usepackage{amstext}
\usepackage{amsthm} % Necessary for proof environment
\usepackage{latexsym}
\usepackage{IEEEtrantools} % provides improved version of eqnarray, called IEEEeqnarray
\usepackage{bm} % provides \bm command for bold math symbols
\usepackage{hyperref} % enables intra-pdf hyperlinks and PDF bookmarks

\makeatletter
\newcommand\RedeclareMathOperator{%
  \@ifstar{\def\rmo@s{m}\rmo@redeclare}{\def\rmo@s{o}\rmo@redeclare}%
}
\newcommand\rmo@redeclare[2]{%
  \begingroup \escapechar\m@ne\xdef\@gtempa{{\string#1}}\endgroup
  \expandafter\@ifundefined\@gtempa
     {\@latex@error{\noexpand#1undefined}\@ehc}%
     \relax
  \expandafter\rmo@declmathop\rmo@s{#1}{#2}}
\newcommand\rmo@declmathop[3]{%
  \DeclareRobustCommand{#2}{\qopname\newmcodes@#1{#3}}%
}
\@onlypreamble\RedeclareMathOperator
\makeatother

\RedeclareMathOperator{\Re}{Re}
\RedeclareMathOperator{\Im}{Im}

\newcommand\vb[1]{\mathbf{#1}}

\begin{document}

%===========================
%   Title, Author, Affiliation
%===========================

\title{Collective terahertz fluctuation modes in a polariton laser}

\author{M.~Em.~\surname{Spotnitz}}
\affiliation{Department of Physics, The University of Arizona, Tucson, AZ 85721}
\affiliation{Wyant College of Optical Sciences, The University of Arizona, Tucson, AZ 85721}

\author{N.H.~\surname{Kwong}}
\affiliation{Wyant College of Optical Sciences, The University of Arizona, Tucson, AZ 85721}

\author{R.~\surname{Binder}}
\affiliation{Wyant College of Optical Sciences, The University of Arizona, Tucson, AZ 85721}
\affiliation{Department of Physics, The University of Arizona, Tucson, AZ 85721}

% Activate to display a given date or no date (if empty), otherwise the current date is printed
\date{January 31, 2023}

%======================================================
%   Abstract   PRL limit 600 characters
\begin{abstract}

 A polariton Bardeen-Cooper-Schrieffer (BCS) state in a semiconductor microcavity is an example of symmetry-broken
states in open systems.
Fluctuations of the order parameter are an important tool to characterize such a state.
With the condensate formed by composite particles,
the set of zero-momentum fluctuations spans an infinite-dimensional electron-hole mode subspace.
We show that collective fluctuation modes with
orbital angular momentum different from that of the order parameter can be obtained with terahertz radiation,
and that a physical manifestation of such modes, which are not Higgs modes, can be terahertz gain.
\end{abstract}
%======================================================

\maketitle

\section{Introduction}

Spontaneous symmetry breaking and macroscopic quantum states are important concepts in physics
\cite{strocchi.2008,beekman-etal.2019}.
One system in which these phenomena are well established are exciton polaritons in
semiconductor microcavities
(see, e.g.,
%THIS IS ORDERED SO THAT THE BEC PAPERS ARE TOGETHER AND THE OUT-COMMENTED ARE REMOVED
\cite{%
fan-etal.97pra,%
cao-etal.97,%
kuwata-gonokami-etal.97,%
duer-etal.1997,%
kira-etal.99b,%
baumberg-etal.2000,%
ciuti-etal.00,%
savvidis-etal.00,%
kwong-etal.01prl,%
keeling_collective_2007,%
schumacher-etal.07prb,%
amo-etal.2009superfluid,%
kamide-ogawa.10,%
liu-etal.15,%
schulze-etal.14,%
kamandardezfouli-etal.14,%
carcamo-etal.20,%
ozturk-etal.2021,%
moskalenko-snoke.00,%
balili-etal.07,%
bajoni-etal.08,%
berney-etal.08,%
amo-etal.2009BEC,%
semkat-etal.09,%
deng-etal.10,%
menard-etal.14,%
schmutzler-etal.15,%
barachati-etal.2018,%
bao-etal.19}),
where polaritonic and photonic Bose-Einstein condensation
\cite{%
moskalenko-snoke.00,%
balili-etal.07,%
bajoni-etal.08,%
berney-etal.08,%
amo-etal.2009BEC,%
semkat-etal.09,%
deng-etal.10,%
menard-etal.14,%
schmutzler-etal.15,%
barachati-etal.2018,%
bao-etal.19,%
pieczarka-etal.2022prb,%
pieczarka-etal.2022},
and polaritonic Bardeen-Cooper-Schrieffer (BCS) states
\cite{%
comte-nozieres.82,%
keeling-etal.05,%
kremp-etal.08,%
kamide-ogawa.10,%
byrnes-etal.10,%
combescot-shiau.15,%
hu-liu.20,%
hu-etal.21}
have been discussed
\footnote{Recently, excitonic Bose-Einstein condensation in bulk semiconductors has been demonstrated \protect\cite{morita-etal.2022}.}.
Here, the condensate wave function, or order parameter, is associated with optically active polaritons, which facilitates its
observation through optical experiments.
These systems are open, dissipative and pumped, hence the physics of these symmetry-broken states can be quite different from their counterparts in thermal equilibrium. For example, the polaritonic order parameter oscillates
 typically at a frequency in the visible or near-infrared spectrum (close to the exciton frequency in the non-condensed state),
 and most studies of
 symmetry-broken states in
 polaritonic systems have characterized the properties of these states using optical probes nearly resonant with the frequency of the order parameter.

However, interesting and potentially new physical effects of the macroscopic quantum state can also be obtained from light fields far detuned from that resonance, for example from terahertz (THz) fields. An example utilizing THz radiation to elucidate the physics of a polaritonic BEC  was given in Ref.\
 \cite{menard-etal.14}.

 Systematic studies of the fluctuation modes of a many-particle system, either condensed or in the normal state, usually involve the system's linear response, triggered by physical fluctuations or weak external probes.
  For the broad class of condensed systems with complex order parameters, where symmetries of the phase(s) are broken, much attention has been paid to the Goldstone (phase) modes and, when they exist \footnote{The conditions for the existence of the Higgs modes are
  	discussed in \cite{varma.2002}.}, the Higgs (amplitude) modes
  \cite{varma.2002,combescot-etal.2006pra,pekker-varma.15,brierley-etal.11,behrle-etal.18,steger-etal.2013,schwarz-etal.2020}.
   Polaritonic condensates involve composite particles with a large number of internal degrees of freedom (the relative motion of electrons and holes making up the excitonic polarization, which is  coupled to the light field in the cavity). This creates a rich landscape of possible fluctuation states. If the broken symmetry is U(1), the only fluctuation mode that can be expected is a simple phase mode.
   In general, all other modes
   require a more detailed classification, not only in terms of phase and amplitude fluctuations, as is usually done, but also electron-hole density fluctuations, as we do here.
The density considered here receives contributions from both the order parameter amplitude and the incoherent pumped reservoir.

 In this study, we investigate the physics of fluctuation modes
 of a polariton laser in the BCS regime triggered by a THz probe.
 Extending our work in Ref.\ \onlinecite{binder-kwong.2021} to the THz case and
 using a many-particle approach based on the diagonalization of the fluctuation matrix, including the electron-hole Coulomb interaction,
 we  obtain all fluctuation modes induced by THz radiation and compare them with those resulting from an `optical' (nearly resonant with the order parameter) probe.
 We find that the orbital angular momentum of the THz-induced fluctuation modes is different from that of the order parameter and that of the conventional optical fluctuation modes.
Both cases, THz and optical, include collective (discrete) modes in addition to the spectral continua. The continuum THz-induced modes can yield THz gain, similar to the case without Coulomb interaction, which we studied in Ref. \onlinecite{spotnitz-etal.2021}. But due to the many-particle Coulomb interactions, we also find collective (discrete) THz-induced fluctuation modes (we label them as `T' modes). We provide a detailed characterization of the physics of these modes, including phase, amplitude and density fluctuations for each degree of freedom.
 Importantly, we find that the new THz-induced collective fluctuation modes yield THz gain which could make them of broad interest in future applications.

Our semiclassical theoretical approach to the THz-induced fluctuation modes of a polariton laser operating in the BCS regime is an extension of previous work
 \cite{binder-kwong.2021}
 that was restricted to probe fields (or fluctuations) with frequencies in the vicinity of the frequency of the order parameter, which in the polariton comprises the interband polarization and the light field in the cavity.
 The dynamical variables are the interband polarization $p(\mathbf{k})$ (where $\mathbf{k}$ is the electronic wave vector),
 the carrier distribution $f(\mathbf{k})$ (same for electrons and holes since we use equal electron and hole masses and relaxation rates),
 and the single-mode cavity laser field $\mathbf{E}_\ell$.
 We use the Hartree-Fock (HF) approximation for $p(\mathbf{k})$ and  $f(\mathbf{k})$, reducing their equations of motion to the semiconductor Bloch equations \cite{chow-etal.94}, amended by phenomenological dephasing, intraband relaxation, non-radiative recombination, and incoherent population pump terms.
 The detailed equations are given in
 section \ref{sec:zeroth-order-theory}.
  Below threshold, this theory yields the lower and upper polariton in the optical response. Above threshold and in steady state, the order parameter $(p^{(0)}(\mathbf{k}),E^{(0)}_\ell)$
  oscillates at the laser frequency $\hbar \omega_\ell$, which is approximately at the fundamental bandgap $E_g$ ($\sim 1.5$eV in GaAs).
  We denote frequencies similar to $E_g$ as interband frequencies, and distinguish them from THz frequencies, which we call intraband frequencies,
  where interband (intraband) refers to the electronic excitations caused by fields oscillating at the corresponding frequencies.
  The complex order parameter has an arbitrary phase factor $e^{i \phi}$, which is fixed in any given realization (spontaneous symmetry breaking).

Fluctuation modes of the laser can be triggered by weak optical-frequency (interband) or THz (intraband) probes.
In Sec.\ \ref{sec:first-order-theory},
the evolution equations of the variables
 $(p(\textbf{k}),f(\textbf{k}),E)$
 are expanded to first order in the probe around the steady-state laser solution. These linear response equations are simplified by angular momentum selection rules: an optical-frequency probe drives fluctuations in the sector with orbital angular momentum $m=0$, and a THz probe drives the modes with $m=\pm 1$.
As detailed in Sec.\ \ref{sec:modedecomp},
 the equations are discretized on a radial-k grid and solved by diagonalization of the fluctuation matrix.

 The benefit of this microscopic approach is that we obtain all fluctuation modes from a many-particle theory within the HF approximation, including discrete (collective) modes and continuous spectra,
 without making any assumptions on the physics of the fluctuation modes.
  The THz response (the induced intraband current) is constructed as an expansion in the eigenmodes of the fluctuation matrix.
  This, in turn, allows us in Sec.\ \ref{sec:results}
  to identify spectral features in the intraband conductivity, and thus the THz absorption spectrum, and to associate these features with specific fluctuation modes.

In Sec.\ \ref{sec:modedecomp}, we formulate the decomposition of fluctuation modes into phase, amplitude, and density oscillation components.
In Secs.\ \ref{sec:phasampresp} and \ref{subsec:mode-characterization},
 we perform an in-depth analysis of the modes'
 oscillation characteristics, and, for most modes, distinguish them from pure phase (Goldstone) or amplitude (Higgs) modes.

\section{Microscopic formalism of electron, hole, and photon dynamics}
\label{sec:zeroth-order-theory}

In this section, we write down the microscopic Hamiltonian for the system of conduction band electrons, valence band holes, and cavity photons, and the equations of motion of field expectation values as used in this paper.
We use bold letters to denote position vectors and physical quantities whose directions are defined in physical space, e.g., wave vector $\mathbf{k}$, electric field $\mathbf{E}$.
An overhead arrow is used to denote a finite array of numbers arranged in column vector form.
For two column vectors, $\vec{a} = (a_1, a_2 , \cdots , a_N)^T$, $\vec{b} = (b_1, b_2 , \cdots , b_N)^T$, the symbol $\vec{a}^{\, T} \vec{b}$ denotes a dot product: $\vec{a}^{\, T} \vec{b} = \sum^{N}_{i=1} a_i b_i$.

\subsection{The Hamiltonian}
\label{subsec:H}

The setup that we consider is an optical microcavity containing a zero-width quantum well.
The fundamental resonance frequency of the cavity is close to the quantum well's band gap.
The system is incoherently pumped to sustain steady state lasing.
The laser's fluctuation spectrum is probed with the linear response to a weak THz field.
A coordinate system is set up in which the $z$ axis is normal to the quantum well's plane.
Our model Hamiltonian for the electrons, holes, and cavity photons is
\begin{widetext}
\begin{align}
	\hat{H} &= \sum_{\alpha, \vb{k}} \varepsilon_{\alpha \vb{k}} a_{\alpha \vb{k}}^{\dagger} a_{\alpha \vb{k}}
	+ \sum_{\lambda \vb{q}} \hbar \omega_{\lambda\vb{q}} c_{\lambda \vb{q}}^{\dagger} c_{\lambda \vb{q}}
	- \frac {1} {\sqrt{\mathcal{A}}} \sum_{\lambda e h \vb{q}, \vb{k}}
	\left[ \Gamma_{eh}^{\lambda} (\vb{k} ,\vb{q}) c_{\lambda \vb{q}} a_{e, \mathbf{k}}^{\dagger}
	a_{h, \mathbf{q}-\mathbf{k}}^{\dagger} + \mathrm{H.c.} \right]
	\label{eq:modelH}\\
	& \quad + \sum_{\nu \alpha \vb{q}, \vb{k}} g_{\alpha}^{\nu} ( \vb{k} + \tfrac{1}{2} \vb{q} )
	A_{T\nu} ( \vb{q} , t ) a_{\alpha , \mathbf{q}+\mathbf{k}}^{\dagger} a_{\alpha , \mathbf{k}}
	+ \frac{1}{2\mathcal{A}} \sum_{\mathbf{k}, \mathbf{k}', \mathbf{q}' \neq 0} \sum_{\mu, \mu'}
	V_{\mathbf{q}'}^{c} a_{\mu, \mathbf{k}}^{\dagger} a_{\mu', \mathbf{k}'}^{\dagger}
	a_{\mu', \mathbf{k}' + \mathbf{q}'} a_{\mu, \mathbf{k} - \mathbf{q}'} .  \nonumber
\end{align}
\end{widetext}
where $a_{e \vb{k}}$, $a_{h \vb{k}}$, and $c_{\lambda \vb{q}}$ are the annihilation operators
for conduction band electrons, valence band holes, and cavity photons respectively.
$\vb{k}$, $\vb{k}'$, $\vb{q}$ and $\vb{q}'$ are 2D wavevectors parallel to the quantum well's plane
(all wavevectors in this paper are in-plane unless specified otherwise), $\lambda$ labels the cavity photon spin, and $\mathcal{A}$ is the normalization area in the plane.
We consider interband transitions only between the highest heavy-hole valence band and the lowest
conduction band. So the band subscripts label the degenerate spin orbitals: $e = \pm 1/2, h = \pm 3/2$.
The subscript $\alpha$ runs through both electron and hole bands.
The subscripts $\mu$ and $\mu^{\prime}$ in the Coulomb interaction term run over the degenerate conduction and valence band spin orbitals, $\mathrm{c} = \pm \frac{1}{2}$ and $\mathrm{v} = \pm \frac{3}{2}$, respectively: $\mu, \mu^{\prime} \in \{\mathrm{c}, \mathrm{v}\}$.
Parabolic bands are used for the charges: $\varepsilon_{e \vb{k}} = \frac {\hbar^2 k^2} {2 m_e} + E_g$ and
$\varepsilon_{h \vb{k}} = \frac {\hbar^2 k^2} {2 m_h}$, where $m_\alpha$ is the effective mass in band $\alpha$
(both $m_e$ and $m_h$ are positive on our case),
and $E_g$ is the band gap. $\omega_{\lambda \vb{q}}$ is the cavity resonance frequency.
The interband eh-laser interaction term is calculated in the rotating wave approximation.
Though it is treated as an input parameter in the numerical calculations,
the interband coupling strength
$\Gamma_{eh}^{\lambda} (\vb{k} ,\vb{q}) $ can be given by
$\Gamma_{eh}^{\lambda} (\vb{k} ,\vb{q}) = \left\vert \vb{d}_{\mathrm{c}\mathrm{v}} \left(\vb{k} \right) \cdot
\bm{\epsilon}_{\ell \lambda} \right\vert \Psi_{\mathrm{cav}} \left(z_{\mathrm{QW}} \right)
\sqrt{2 \pi \hbar \omega_{\lambda \vb{q}} /\epsilon_{b}}$.
The interband dipole moment is
$\vb{d}_{\mathrm{c}\mathrm{v}} \left(\vb{k} \right) = i e \hbar \left\langle\mathrm{c},
\vb{k} \right\vert \hat{\vb{p}}\left\vert \mathrm{v}, \vb{k} \right\rangle / [ m_0 \, \Delta
E_{\mathrm{c v}} (\vb{k}) ]$,
where $m_0$ is the free-space electron mass, $e$ is the magnitude of the electron's charge ($e > 0$),
the states $\left\vert \mathrm{c}, \vb{k} \right\rangle$ ($\left\vert \mathrm{v}, \vb{k} \right\rangle$)
in the electron momentum matrix element are conduction (valence) band Bloch wave functions, and
$\Delta E_{\mathrm{c v}} (\vb{k}) = \varepsilon_{e \vb{k}} + \varepsilon_{h \vb{k}}$.
$\Psi_{\mathrm{cav}} \left(z_{\mathrm{QW}} \right)$ is the cavity photon mode 1D wave function along the $z$
direction evaluated at the position of the quantum well $z_{\mathrm{QW}}$, $\epsilon_{b}$ is the background
dielectric function inside the cavity, and $\bm{\epsilon}_{\ell \lambda}$ is the polarization unit vector of the
optical field.
(Some nuances of the relation between the interband dipole and momentum matrix elements are discussed in
Ref.\ \cite{gu-etal.13}; see also Ref.\ \cite{mahon-etal.19}.)
Conservation of angular momentum in the $e = \pm 1/2$ to $h = \pm 3/2$ transitions requires (see, e.g., Refs.\ \onlinecite{hu-etal.21} \& \onlinecite{spotnitz-etal.2021}) the ``circular selection rules''
$\vb{d}_{\mathrm{c}\mathrm{v}} \left(\vb{k} , \vb{q} \right) \cdot \bm{\epsilon}_{\ell \lambda} =
d_{\mathrm{c}\mathrm{v}} \left(\vb{k} , \vb{q} \right)
\delta_{|e+h|,1}
\delta_{e+h,\lambda}$
where the angular momentum labels in the conduction-valence band picture are related to those in the electron-hole picture via $\mathrm{c}=e$ and $\mathrm{v}=-h$.
That is, for a given conduction band, the corresponding valence band and laser-photon polarization are fixed.

The THz probe is treated in the Hamiltonian Eq.\ (\ref{eq:modelH}) as a classical applied vector potential $\vb{A}_{T} (t)$.
We use a gauge in which the scalar potential is zero \cite{mahon-etal.19} so that $\vb{E}_{T} ( \vb{x} , t ) = - \tfrac{1}{c} \partial \vb{A}_{T} ( \vb{x} , t ) / \partial t$, with $\vb{x}$ being the 3D spatial coordinates.
The probe induces intraband transitions with the coupling $g^{\nu}_{\alpha}(\vb{k})$, where $\nu$ labels the polarization state of the THz field.
With the approximations of isotropy and small transverse THz wavevector $\vb{q}\ll\vb{k}$, the coupling strength is evaluated as
\begin{equation}
	g^{\nu}_{\alpha} \left(\vb{k} \right) = - \frac{ s_\alpha e }{ m_{\alpha} c} \hbar \vb{k} \cdot \bm{\epsilon}_{T \nu},
	\label{eq:gnuform}
\end{equation}
where $\bm{\epsilon}_{T \nu}$ is the polarization unit vector for the THz field and $s_\alpha$ is the sign of the particle's charge: $s_e = -1 , s_h = 1$.
	The intraband e/h-EM interaction operator of Eq.\ \eqref{eq:modelH} states that intraband electronic transitions are possible for $\mathbf{q}=0$ if $\mathbf{k} \neq 0$.
	Angular momentum is conserved between an electron and a THz photon by the factor of $\hbar\mathbf{k}$ in Eq.\ \eqref{eq:gnuform}, which changes the angular quantum number $m$ by $\pm 1$ for the electronic orbital motion.
	Energy is conserved in these intraband transitions by changes in an excitonic state, or as described in this paper, by transitions the band and its light-induced counterpart (cf. Fig. \ref{fig:bandstruct} below for more details).
	This interaction does not change the electron or hole spin, of which its strength is independent; and due to the assumption of electronic isotropy, its strength is independent of the THz field polarization.
The quasi-2D Coulomb interaction energy is
\begin{equation}
	V_{\mathbf{q}}^{c} \equiv \frac{2\pi e^{2}}{\epsilon_{b} } \frac{1}{\left|\mathbf{q}\right| + \kappa_{0}} . \label{eq:Vqcdef}
\end{equation}
where $\kappa_{0}$ is a small, constant screening wavenumber.

Because typical THz wavenumbers $q$ in a dielectric are much less than a typical electron quasimomentum $k$, $q \ll k$, it was previously found in Ref.\ \cite{spotnitz-etal.2021} that in calculating the THz response (specifically the conductivity $\sigma_{T}$) of the QW it is a very good approximation to take the in-plane THz field wavevector to be zero, $\vb{q}_{\|} \approx 0$. Varying the angle of incidence of the THz probe does affect its transmissivity, as was shown in Fig.\ 12 in Ref.\ \cite{spotnitz-etal.2021}. However, this is almost entirely due to Maxwell's equations, and their specific results in Eqs.\ (A12)--(A16) of Ref.\ \cite{spotnitz-etal.2021}, and not because of any change in the THz-induced conductivity $\sigma_{T}$. In this paper, we again use Eqs.\ (A12)--(A16) of Ref.\ \cite{spotnitz-etal.2021}, and the difference from Ref.\ \cite{spotnitz-etal.2021} arises in the calculation of $\sigma_{T}$.
Therefore, the effects of the probe geometry on the THz transmissivity have already been adequately shown in Fig.\ 12 of Ref.\ \cite{spotnitz-etal.2021}, and it is no longer necessary to consider the effects of changing the THz angle of incidence. Instead, we simplify the math by taking the THz probe $\vb{E}_{T}$ to be normally incident:
\begin{equation}
	\mathbf{E}_{T} (z,t) \cdot \bm{\epsilon}_{\nu} \equiv E_{T\nu}(z,t)
	= \int \frac{\mathrm{d}\omega}{2\pi} e^{i(q_{z}z - \omega t)} E_{T\nu}(\omega)
\end{equation}
where
$q_{z} = \sqrt{\epsilon_{b}} \omega/c$ and $\bm{\epsilon}_{\nu}$ is the polarization unit vector, equal to $\hat{y}$ for $\nu =y$ or $\hat{x}$ for $\nu = x$.
The quantum well microcavity's response to this probe consists of fluctuations with zero (in-plane) momentum (the steady lasing state before the THz probe arrives being isotropic in the plane). Accordingly, only the $\vb{q}= \vb{0}$ part of $\Gamma_{eh}^{\lambda} (\vb{k} ,\vb{q})$ is in use. We approximate this relevant part, $\Gamma_{eh}^{\lambda} (\vb{k} ,\vb{0})$, by a function, denoted by $\Gamma_{eh}^{\lambda} (\vb{k})$, that equals a constant for $|\mathbf{k}|$ less than a set value $k_{max}$ and zero for $|\mathbf{k}| > k_{max}$. The numerical value of the constant is adjusted such that the splitting between the exciton and the lower polariton (LP) resonance, sometimes called vacuum Rabi splitting and denoted by $\Omega_R$, has a given value consistent with state of the art microcavities.
We also use the notation $\omega_{\lambda \vb{0}} \equiv \omega^{\lambda}_{cav}$.

%===========================
%   Equations of Motion
%===========================
\subsection{Equations of motion of the cavity field, the interband polarization, and the charge densities}

Using the Hamiltonian in Eq.\ \eqref{eq:modelH}, single-time
equations of motion are derived for the slowly-varying envelope of the electron-hole polarization at zero
center-of-mass momentum,
$ p_{e h} (\vb{k}, t ) \equiv \left\langle a_{h, -\vb{k}} (t) a_{e, \vb{k}} (t)
\right\rangle e^{i \omega_{\ell} t}$, the occupation functions $f_{\alpha} (\vb{k}, t )
\equiv \left\langle a_{\alpha, \vb{k}}^{\dagger} (t) a_{\alpha, \vb{k}}  (t) \right\rangle$,
where $\alpha \in \{e,h\}$, and the envelope of the laser field amplitude (the squared
magnitude of which is the 2D photon density in the designated mode)
$E_{\ell \lambda} ( t ) \equiv (1/\sqrt{\mathcal{A}})\left\langle c_{\lambda, \vb{q}=\vb{0}}  (t)
\right\rangle e^{i \omega_{\ell} t}$. $\omega_{\ell}$ is the laser frequency, which is obtained
from solving the steady state equations absent the THz probe.
The circular selection rules require that $p_{e h} (\vb{k}, t ) = 0$ for $(e,h) \notin \{ (\frac{1}{2},-\frac{3}{2}), (-\frac{1}{2},\frac{3}{2})\}$.
In an equation of motion for $f_{e}$, the value for the index $h$ in factors of $p_{eh}$ should be chosen so that $(e,h) = (\frac{1}{2},-\frac{3}{2})$ or $=(-\frac{1}{2},\frac{3}{2})$.
The many-body dynamics is treated at the same approximate level as the semiconductor Bloch equation (SBE).
Photonic correlations, involving non-factorizable parts of expectation values of products of photon operators or products of photon and charge operators, are ignored.
Effects of Coulomb correlations beyond the SBE are modeled by appropriate relaxation and dephasing terms.
The Hamiltonian Eq.\ \eqref{eq:modelH} does not account for pumping, cavity loss via emission of the laser field,
or interactions with the environment, e.g., phonons.
These effects are also modeled phenomenologically,
in the same way as was done in Refs.\ \onlinecite{hu-etal.21,spotnitz-etal.2021,binder-kwong.2021}.
We set $m_{e} = m_{h}$ and also assume all the incoherent relaxation rates are the same for the electrons and the holes. These conditions imply $f_{e=-1/2}(\vb{k},t) = f_{h=3/2}(-\vb{k},t)$ and $f_{e=1/2}(\vb{k},t) = f_{h=-3/2}(-\vb{k},t)$. We keep the electron distribution $f_e (\vb{k} , t)$ as the dynamical variable for each spin configuration. This electron-hole symmetry assumption is made for simplicity. Reverting to the correct mass ratio and unequal relaxation rates would not change any physical conclusions in the paper.
Under these approximations, we obtain the following equations of motion,
\begin{widetext}
\begin{multline}
	i\hbar \frac{\partial }{\partial t} p_{e h} (\mathbf{k}, t) = \left( \frac{\hbar^{2}k^{2}}{2m_{r}}+E_{g}-\frac{2}{\mathcal{A}}\sum\limits_{\mathbf{k}^{\prime }}V_{\mathbf{k-k}^{\prime }}^{c}f_e(\mathbf{k}^{\prime}, t) -\hbar \omega_{\ell}-i\gamma + \sum_\nu g^{\nu}(\mathbf{k}) A_{T \nu}(t) \right) p_{e h} (\mathbf{k},t)  \\
	-\left[ 1-2f_e(\mathbf{k},t)\right] \left[ \sum_{\lambda} \Gamma^{\lambda}_{e h} (\vb{k}) E_{\ell \lambda} (t) +\frac{1}{\mathcal{A}}\sum\limits_{\mathbf{k}^{\prime }}V_{\mathbf{k-k}^{\prime }}^{c} p_{e h} (\mathbf{k}^{\prime},t)\right] , \label{eq:SBE-P-dot}
\end{multline}
where $g^\nu (\vb{k}) \equiv g^\nu_{e}(\vb{k}) + g^\nu_{h}(-\vb{k}) = \frac{e}{m_{r}c} \hbar \vb{k} \cdot \bm{\epsilon}_{T \nu}$, with the reduced mass $m_{r}$ given by $\frac{1}{m_{r}} = \frac{1}{m_{e}} + \frac{1}{m_{h}}$,
\begin{multline}
	\hbar \frac{\partial}{\partial t} f_e (\vb{k},t) = 2 \Im \left\{  \left[ \sum_{\lambda} \Gamma^{\lambda \ast}_{e h} (\vb{k}) E_{\ell \lambda}^{\ast} (t) + \frac{1}{\mathcal{A}} \sum_{\vb{k}^{\prime}} V_{\vb{k}-\vb{k}^{\prime}}^{c} p^{\ast}_{e h} (\vb{k}^{\prime},t)  \right] p_{e h}(\vb{k},t)\right\} \\
	- \gamma_{F} \left[f_e(\vb{k},t) - f_{F}(\vb{k},t)\right] - \gamma_{nr} f_e(\vb{k},t)- \gamma_{p} \left[f_e(\vb{k},t) - f_{p}(\vb{k})\right] ;
	\label{eq:fmot}
\end{multline}
and %
\footnote{An alternative to the single-mode equation for the cavity field based on the
	propagation of the light through the entire microcavity structure has been
	given in Ref.\ \protect\onlinecite{carcamo-etal.20}.}
\begin{equation}
	i \hbar \frac{\partial}{\partial t} E_{\ell \lambda} (t)
	= (\hbar \omega^{\lambda}_{cav} - \hbar\omega_{\ell} - i \gamma_{cav}) E_{\ell \lambda} (t) - \frac{N_{QW}}{\mathcal{A}} \sum_{ \vb{k}eh} \Gamma^{\lambda \ast}_{e h} (\vb{k}) p_{e h}(\vb{k},t).
	\yesnumber \label{eq:Esingmod}
\end{equation}
\end{widetext}

The approximation of the incoherent scattering terms in terms of relaxation rates has been previously described in detail in Eqs.\ (B1)--(B8), (B16), and (B17) of \cite{spotnitz-etal.2021}; as well as in Eqs.\ (5)--(8) of the supplemental material for \cite{binder-kwong.2021}, and in Eqs.\ (23)--(26) of \cite{hu-etal.21}.
In Eqs.\ \eqref{eq:SBE-P-dot}, \eqref{eq:Esingmod}, and below, $\gamma$ is the dephasing rate of e-h pairs, $\gamma_{cav}$ is the decay rate of the cavity field, and $N_{QW}$ is the number of quantum wells. In Eq.\ \eqref{eq:fmot}, $\gamma_{F}$ is the thermalization rate, $\gamma_{p}$ is the incoherent pump rate, and $\gamma_{nr}$ is the non-radiative decay rate.
We define the total distribution relaxation rate $\gamma_{f}\equiv \gamma _{F}+\gamma _{p}+\gamma _{nr}$.
Incoherent pumping is modeled by the term $- \gamma_{p} \left[f_e(\vb{k},t) - f_{p}(\vb{k})\right]$, which drives the distribution function $f_e(\vb{k},t)$ towards a pump-induced Fermi function $f_{p}(\vb{k})$ at the rate $\gamma_{p}$.
The pump chemical potential $\mu_{p}$ is chosen such that the density $n_{p} = 2 \int \frac{\mathrm{d}^{2} k}{(2\pi)^2} f_{p}(\vb{k})$, corresponds to the chosen pump density, which is an input parameter in this theory.
Intraband carrier-carrier scattering drives the distribution functions towards Fermi functions at the rate $\gamma_{F}$, without changing the total carrier density $n(t)$ in each band. This is included via the term $- \gamma_{F} \left[f_e(\vb{k},t) - f_{F}(\vb{k},t)\right]$.
The thermal chemical potential $\mu_{F}$ is chosen at each time such that $n(t) =2\int \frac{d^{2}k}{(2\pi )^{2}}f_e(\mathbf{k},t)=2\int \frac{d^{2}k}{(2\pi)^{2}}f_{F}(\mathbf{k},t)   .$
$f_{F}(\vb{k})$ and $f_{p}(\vb{k})$ are Fermi distributions with distinct chemical potentials:
\begin{equation}
	f_{x} \left(\vb{k}; \mu_{x}\right) = \frac{1}{e^{\left( \varepsilon_{\vb{k}} - \mu_{x} \right)/k_B T} + 1} , \label{fermidisteq}
\end{equation}
where $x \in \{F, p\}$, and $T$ is an effective e-h temperature, also set as a parameter.

%===========================
%   Zeroth order equations
%===========================
\section{Linear response to a Terahertz probe}
\label{sec:first-order-theory}

\subsection{Laser Steady State}

The stationary solutions are given by taking the $i \hbar \frac{\partial}{\partial t}$ terms and $A_{T \nu}$ to be zero in Eqs.\ \eqref{eq:SBE-P-dot}--\eqref{eq:Esingmod}.
They are denoted with a superscript $(0)$: $f^{(0)}_e (\vb{k})$, $p^{(0)}_{e h} (\vb{k})$ and $E^{(0)}_{\ell \lambda}$. The use of $0$ references that the stationary solutions are zeroth-order in the perturbing THz field $A_{T \nu}(t)$.
We limit ourselves to $s$-wave solutions, meaning that all $\mathbf{k}$-dependent functions depend only on the magnitude of the wave vector, $k=|\mathbf{k}|$. One steady state solution, with $p^{(0)}_{e h} (\vb{k}) = 0$ and $E^{(0)}_{\ell \lambda} = 0$, represents the non-lasing, `normal' state. When the pump density $n_p$ is raised above a threshold, additional solutions, representing the lasing state, with non-zero $p^{(0)}_{e h} (\vb{k})$, $E^{(0)}_{\ell \lambda}$, and $\omega_{\ell}$, appear.
Explicit expressions of these solutions can be found as Eqs.\ (9)--(14) in the Supplement to \cite{binder-kwong.2021}, where the stationary frequency is denoted in that paper by $\omega_{0}$ and in this paper by $\omega_{\ell}$.
In practice, we obtain the laser solutions numerically by solving Eqs.\ \eqref{eq:SBE-P-dot}--\eqref{eq:Esingmod} (with $A_{T \nu} = 0$) in time, using a small fluctuation to trigger the normal-to-lasing phase transition, and evolving the solution to a steady state.

The laser solution spontaneously breaks a $U(1)$ symmetry. The overall phase of the set of complex variables $(p^{(0)}_{e h} (\vb{k}), E^{(0)}_{\ell \lambda})$ is not determined by the equations, and there are infinitely many solutions which are assigned different values of this overall phase but are otherwise equivalent.
Formally, one can generalize the concept of a gap function $\Delta (\mathbf{k})$ to the polariton lasing or polariton BCS-like state (where BCS stands for Bardeen-Cooper-Schrieffer), which alternatively can be referred to as the effective Rabi frequency $\Omega _{eff}(\mathbf{k}) $,
\begin{equation}
	\Delta (\mathbf{k}) \equiv \Omega _{eff}(\mathbf{k})\
	= \Gamma^{\lambda}_{e h} (\vb{k}) E_{\ell \lambda}^{(0)} +\frac{1}{\mathcal{A}}\sum\limits_{\mathbf{k}^{\prime }}V_{\mathbf{k-k}%
		^{\prime }}^{c} p^{(0)}_{e h} (\mathbf{k}^{\prime})     .
	\label{eq:definition-BCS-gap-and-eff-Rabi-frequ}
\end{equation}
The original $T=0$K BCS state for superconductors, which follows from a HF theory for Cooper pairs in an interacting Fermi gas is given, for example, on pp.\ 326-336 of Ref.\ \cite{fetter-walecka.71}.
An analogous theory for polaritons was formulated in Refs.\ \cite{byrnes-etal.10,kamide-ogawa.10}.

The quasi-phenomenological approach for the BCS-like gap in the polariton laser, as previously formulated in Refs.\ \onlinecite{hu-etal.21,binder-kwong.2021,spotnitz-etal.2021}, is further corroborated in this paper.
The expressions used there are formally the same as in the standard BCS theory.
We reproduce them here for convenience in the notation of Ref.\ \onlinecite{binder-kwong.2021}:
\begin{equation}
	\tilde{\xi}(\mathbf{k}) = \frac{\hbar^{2}k^{2}}{2m_{e}} + \Sigma_{e}^{HF}(\mathbf{k})-\frac{1}{2}\left( \hbar \omega_{\ell}-E_{g}\right)
	\label{eq:xiergdef}
\end{equation}
where the single-particle Hartree-Fock self-energy is
\begin{equation}
	\Sigma_{e}^{HF}(\mathbf{k})= - \frac{1}{\mathcal{A}}\sum\limits_{\mathbf{k}^{\prime }}V_{%
		\mathbf{k-k}^{\prime }}^{c}f^{(0)}_{e}(\mathbf{k}^{\prime }),
\end{equation}
and the excitation energies are
\begin{equation}
	\tilde{E}(\mathbf{k})=\sqrt{\tilde{\xi }^{2}(\mathbf{k})+|\Delta (%
		\mathbf{k})|^{2}}  . \label{equ:E-single-particle-open}
\end{equation}

\begin{figure}
	\centering
	\includegraphics[width=3in]{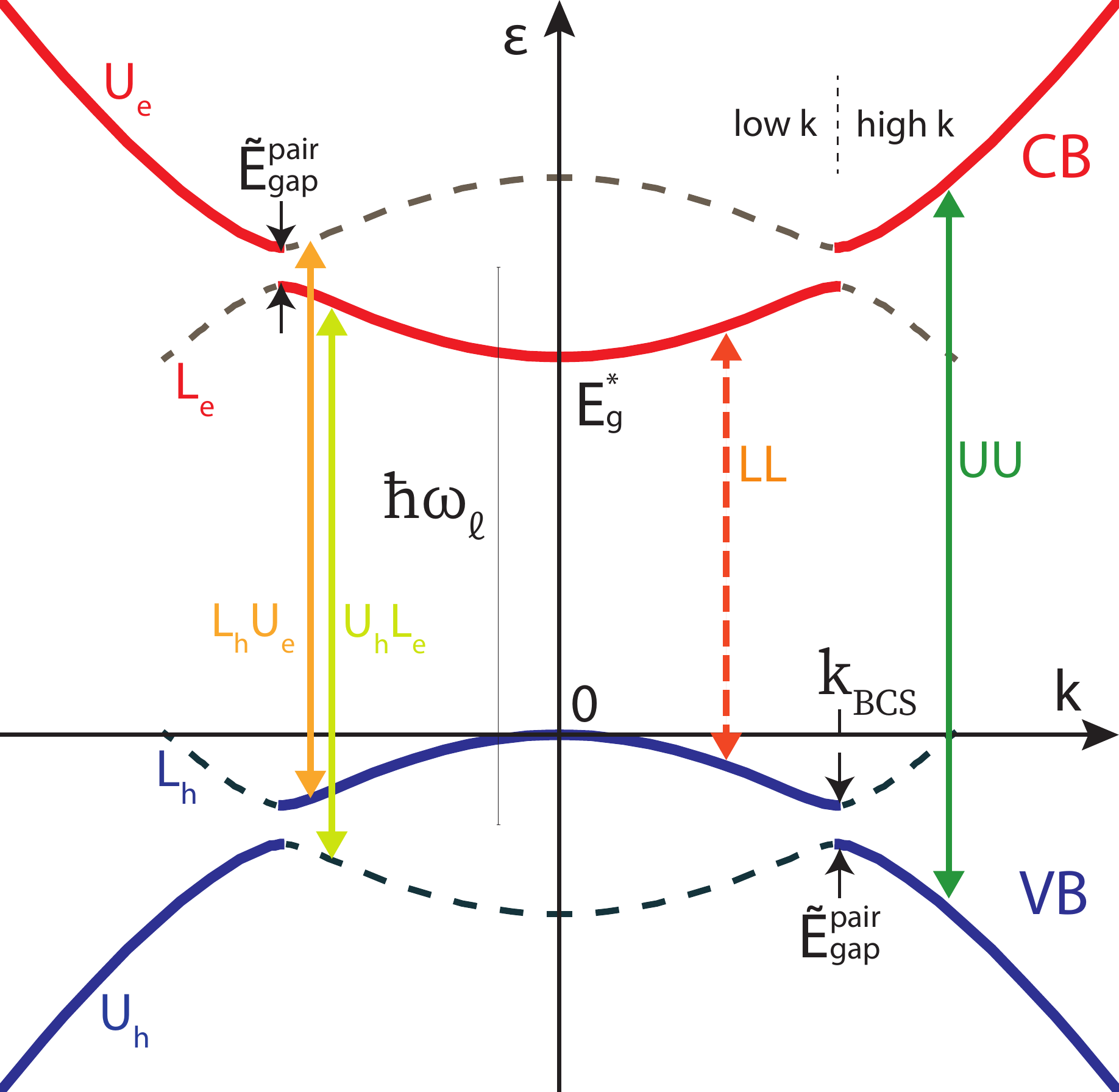}
	\caption{
		(Color online.)
		Schematic of the parabolic two-band band structure
		with the conduction band (CB) in red and the valence band (VB) in blue, separated by the HF Coulomb-renormalized bandgap $E_{g}^{\ast}$, and further renormalized by the electron-hole interaction and the light field (dressed bands), cf. spectral function in \cite{kremp-etal.08,yamaguchi-etal.15}.
		The BCS renormalization induces a gap with magnitude $\tilde{E}^{pair}_{gap}$, and separates the CB (VB) into U$_{e}$ and L$_{e}$ (U$_{h}$ and L$_{h}$, in the hole picture) branches.
	UU, LL, LU and UL transitions are indicated (the latter two resulting in the decay continuum).
	Dashed lines denote smaller, but nonzero, spectral weight for the branch.
		All vertical transitions between the dressed bands (both solid and dashed lines) are possible.
	The THz-frequency transitions result from those indicated by the subsequent subtraction of the frequency $\hbar\omega_{\ell}$ through coherent mixing.
}
\label{fig:bandstruct}
\end{figure}

It is useful to formulate the theory in terms of pair energies, rather than single-particle (electron or hole) energies.
This has been done in Ref.\ \cite{comte-nozieres.82}, where the HF theory was formulated in terms of pair energies which are twice the single-particle energies,
$\xi ^{pair}(\mathbf{k})=2\xi (\mathbf{k})$, and similarly the pair excitation energy
is $\tilde{E}^{pair}(\mathbf{k}) = 2 \tilde{E}(\mathbf{k})$.
We then obtain the pair BCS-like gap from minimizing $\tilde{E}(\mathbf{k})$
\begin{equation}
	\tilde{E}_{gap}^{pair} = 2 \min_{k} \tilde{E}(\mathbf{k}), \label{eq:def-epairgap}
\end{equation}
As will be shown in section \ref{sec:modedecomp}, the linear THz response can be formulated as an eigenvalue problem.
We then show in Fig.\ \ref{fig:eigenergkspect}(a) that this quasi-phenomenological BCS-like approach indeed yields the same pair excitation dispersion relation, $\tilde{E}^{pair}(k)$, as we obtain from the much more rigorous linear response spectrum given by the eigenvalues $\Re \varepsilon_{k}$.
Conversely, the identification of the $\Re \varepsilon_{k}$ UU and LL continua with $\tilde{E}^{pair}(k)$ enables the interpretation of the finite-frequency linear excitations as BCS-like excited pairs.
The UU and LL notation and a schematic plot of the renormalized bands are given in Fig.\ \ref{fig:bandstruct}.

%===========================
%   THz Probe Evolution
%===========================
\subsection{Linear THz response}
% Construction of the THz response -- perturbative equations

After a desired steady state laser solution is prepared as described in the previous subsection, we switch on a small continuous-wave external THz field $\vb{A}_{T}(\omega)$ and calculate the linear response of the laser to this probe.
%Specifically, we allow $A_{T} \neq 0$ in Eq.\ \eqref{eq:SBE-P-dot}.
Specifically, we write the interband polarization, the charge density, and the cavity photon field in the presence of $\vb{A}_{T}$ as $p_{e h}(\mathbf{k},t)=p^{(0)}_{e h}(\mathbf{k})+ p^{(1)}_{e h}(\mathbf{k},t)$, $f_e(\mathbf{k})=f^{(0)}_e(\mathbf{k})+ f^{(1)}_e(\mathbf{k},t)$ and $E_{\ell \lambda}(t)=E_{\ell \lambda}^{(0)} + E_{\ell \lambda}^{(1)}(t)$, where the perturbative quantities of $p^{(1)}_{e h}(\mathbf{k},t)$, $f^{(1)}_e(\mathbf{k},t)$ and $E_{\ell \lambda}^{(1)}(t)$ are taken to be on the order of $\vb{A}_{T}$.
Inserting this form of $p_{e h}(\mathbf{k},t)$, $f_e(\mathbf{k},t)$ and $E_{\ell \lambda} (t)$ in Eqs.\ \eqref{eq:SBE-P-dot}--\eqref{eq:Esingmod} and linearizing around the steady state solution,
%and using the fact that the zeroth-order (in $E_{T}$) quantities $P^{(0)}(\mathbf{k})$, $f^{(0)}(\mathbf{k})$ and $E_{\ell}^{(0)}$\ are stationary solutions,
we obtain the first-order equations for $p^{(1)}_{e h}(\mathbf{k},t)$, $f^{(1)}_e(\mathbf{k},t)$ and $E_{\ell \lambda}^{(1)}(t)$ in the time domain as:
%In the time domain, the linearized equations comprise the linearized
%equation for the interband polarization,
\begin{widetext}
\begin{eqnarray}
	i\hbar \frac{\partial }{\partial t}p^{(1)}_{e h}(\mathbf{k}) &=&\left( \frac{%
		\hbar ^{2}k^{2}}{2m_{r}}+E_{g} +2\Sigma_{e}^{HF}(\mathbf{k})-\hbar \omega
	_{\ell}-i\gamma \right) p^{(1)}_{e h}(\mathbf{k})  \nonumber \\
	&&-\frac{2}{\mathcal{A}} p^{(0)}_{e h}(\mathbf{k})  \sum_{\mathbf{k}^{\prime }}V_{\mathbf{k-k}^{\prime
	}}^{c}f^{(1)}_{e}(\mathbf{k}^{\prime }) \nonumber \\
	&&-\left[ 1-2f^{(0)}_{e}(\mathbf{k})\right] \left[\sum_{\lambda} \Gamma^{\lambda}_{e h} (\vb{k})E_{\ell \lambda}^{(1)}+\frac{1}{\mathcal{A}}\sum\limits_{\mathbf{k}^{\prime }}V_{%
		\mathbf{k-k}^{\prime }}^{c}p^{(1)}_{e h}(\mathbf{k}^{\prime })\right]  \nonumber
	\\
	&&+2f^{(1)}_{e}(\mathbf{k}) \left[ \sum_{\lambda} \Gamma^{\lambda}_{e h} (\vb{k})E_{\ell \lambda}^{(0)}+\frac{1}{\mathcal{A}}%
	\sum\limits_{\mathbf{k}^{\prime }}V_{\mathbf{k-k}^{\prime }}^{c} p^{(0)}_{e h}(\mathbf{k}^{\prime })\right] \nonumber \\
	&& + \sum_{\nu} g^{\nu}(\vb{k}) A_{T \nu} p^{(0)}_{e h}(\vb{k}) \label{eq:delta-P-dot}
\end{eqnarray}%
%the linearized equation for the complex conjugate of the interband
%polarization,
\begin{eqnarray}
	-i\hbar \frac{\partial }{\partial t}p^{(1) \ast}_{e h}(\mathbf{k}) &=&\left(
	\frac{\hbar ^{2}k^{2}}{2m_{r}}+E_{g} +2\Sigma_{e}^{HF}(\mathbf{k})-\hbar
	\omega_{\ell}+i\gamma \right) p^{(1) \ast}_{e h}(\mathbf{k})  \nonumber \\
	&&-\frac{2}{\mathcal{A}} p^{(0) \ast}_{e h}(\mathbf{k})  \sum\limits_{\mathbf{k}^{\prime }}V_{\mathbf{k-k}^{\prime
	}}^{c}f^{(1)}_{e}(\mathbf{k}^{\prime }) \nonumber \\
	&& -\left[ 1-2f^{(0)}_{e}(\mathbf{k})\right] \left[ \sum_{\lambda} \Gamma^{\lambda \ast}_{e h} (\vb{k})E_{\ell \lambda}^{(1)\ast}
	+\frac{1}{\mathcal{A}} \sum\limits_{\mathbf{k}^{\prime }}V_{\mathbf{k-k}^{\prime }}^{c}
	p^{(1) \ast}_{e h}(\mathbf{k}^{\prime })\right]  \nonumber \\
	&&+2f^{(1)}_{e}(\mathbf{k})\left[ \sum_{\lambda} \Gamma^{\lambda \ast}_{e h} (\vb{k})E_{\ell \lambda}^{(0)\ast}+\frac{1}{\mathcal{A}}%
	\sum\limits_{\mathbf{k}^{\prime }}V_{\mathbf{k-k}^{\prime }}^{c}p^{(0) \ast}_{e h}(%
	\mathbf{k}^{\prime })\right] \nonumber \\
	&& + \sum_{\nu} g^{\nu}(\vb{k}) A_{T \nu}^{\ast} p^{(0) \ast}_{e h}(\vb{k})  \label{eq:delta-P-star-dot}
\end{eqnarray}%
%the linearized equation for the distribution function,
\begin{eqnarray}
	i\hbar \frac{\partial }{\partial t}f^{(1)}_{e}(\mathbf{k}) &=& \left[ \sum_{\lambda} \Gamma^{\lambda \ast}_{e h} (\vb{k})E_{\ell \lambda}^{(0)\ast}+\frac{1}{\mathcal{A}}\sum\limits_{\mathbf{k}^{\prime }}V_{%
		\mathbf{k-k}^{\prime }}^{c}p^{(0) \ast}_{e h}(\mathbf{k}^{\prime })\right] p^{(1)}_{e h}(\mathbf{k})  \nonumber \\
	&&+\left[ \sum_{\lambda} \Gamma^{\lambda \ast}_{e h} (\vb{k})E_{\ell \lambda}^{(1)\ast}+\frac{1}{\mathcal{A}}\sum\limits_{\mathbf{k}^{\prime }}V_{\mathbf{k-k%
		}^{\prime }}^{c}p^{(1) \ast}_{e h}(\mathbf{k}^{\prime })\right] p^{(0)}_{e h}(%
	\mathbf{k})  \nonumber \\
	&&-\left[ \sum_{\lambda} \Gamma^{\lambda}_{e h} (\vb{k})E_{\ell \lambda}^{(0)}+\frac{1}{\mathcal{A}}\sum\limits_{\mathbf{k}%
		^{\prime }}V_{\mathbf{k-k}^{\prime }}^{c}p^{(0)}_{e h}(\mathbf{k}^{\prime })\right]
	p^{(1) \ast}_{e h}(\mathbf{k},t)  \nonumber \\
	&& -\left[ \sum_{\lambda} \Gamma^{\lambda}_{e h} (\vb{k})E_{\ell \lambda}^{(1)}+\frac{1}{\mathcal{A}}%
	\sum\limits_{\mathbf{k}^{\prime }}V_{\mathbf{k-k}^{\prime }}^{c}p^{(1)}_{e h}(%
	\mathbf{k}^{\prime })\right] p^{(0) \ast}_{e h}(\mathbf{k}) \nonumber \\
	&&-i \gamma_{f} f^{(1)}_{e}(\mathbf{k})
	\label{eq:delta-f-dot}
\end{eqnarray}%
%the linearized equation for the cavity field,
\begin{equation}
	i\hbar \frac{\partial }{\partial t}E_{\ell \lambda}^{(1)}=\left( \hbar \omega _{cav}-\hbar
	\omega _{\ell}-i\gamma _{cav}\right) E_{\ell \lambda}^{(1)}- \frac{N_{QW}}{\mathcal{A}}%
	\sum\limits_{\mathbf{k}eh}\Gamma^{\lambda \ast}_{e h} (\vb{k})p^{(1)}_{e h}(\mathbf{k},t)
	\label{eq:delta-E-dot}
\end{equation}%
%and the linearized equation for the complex conjugate of the cavity field,
\begin{equation}
	-i\hbar \frac{\partial }{\partial t}E_{\ell \lambda}^{(1)\ast}=\left( \hbar \omega
	_{cav}-\hbar \omega _{\ell}+i\gamma _{cav}\right) E_{\ell \lambda}^{(1)\ast}-\frac{%
		N_{QW}}{\mathcal{A}}\sum\limits_{\mathbf{k}eh}\Gamma^{\lambda}_{e h} (\vb{k})p^{(1) \ast}_{e h}(%
	\mathbf{k})  . \label{eq:delta-E-star-dot}
\end{equation}
\end{widetext}

\subsection{Orbital angular momentum selection rules for THz fluctuations}
\label{subsec:orb-ang-mom-sel}

In this subsection, we expand the linear response equations Eqs.\ \eqref{eq:delta-P-dot} - \eqref{eq:delta-E-star-dot} in an orbital angular momentum basis in $\vb{k}$ space.
Since the Hamiltonian and the steady state are isotropic in the plane, one expects the linear `susceptibility' connecting the fluctuations to the probe to be diagonal in angular momentum space.
For a $\vb{q} = \vb{0}$ plane-wave interband (typically optical-frequency) probe, the angular momentum of the absorbed photon accounts for the `spin' change in the unit cell orbital,
and the coupling to the charges' motion within the band is circularly symmetric
(so-called first-class dipole allowed transitions with $s$-like electron-hole envelope function, or in other words, zero angular momentum associated with the relative electron-hole motion).
For the $\vb{q} = \vb{0}$ plane-wave intraband THz probe considered here, the coupling $\sum_{\nu} g^{\nu}(\vb{k}) A_{T \nu} = \sum_{\nu} \frac{e}{m_{r}c} \hbar \vb{k} \cdot \bm{\epsilon}_{T \nu} A_{T \nu}$ transfers (one unit of) angular momentum between the THz photon and the electron's intraband motion. In short, the interband and intraband probes excite respectively (denoting the in-plane intraband orbital angular momentum by $m$) the $m=0$ and $m=\pm 1$ sectors of the fluctuation modes of the laser.

%The first-order polarization $p^{(1)}_{e h} (\vb{k})$ depends on $g(\vb{k})$, as can be seen in Eqs.\ \eqref{eq:delta-P-dot} and \eqref{eq:delta-P-star-dot}.
%Through their dependence on $p^{(1)}_{e h} (\vb{k})$, as shown in Eqs.\ \eqref{eq:delta-f-dot}--\eqref{eq:delta-E-star-dot}, $f^{(1)}_{e} (\vb{k})$ and $E^{(1)} (\vb{k})$ depend on $g(\vb{k})$ as well.
%For simplicity, the THz polarization is taken in the $y$-direction, $\bm{\epsilon}_{T} = \bm{y}$. Then,
%The THz coupling is $g(\vb{k}) = \frac{e}{m_{r}c} \hbar \vb{k} \cdot \bm{\epsilon}_{T}$. This is an explicit function of the vector components $k_{x}$ and $k_{y}$. Therefore, it is no longer possible to make the ``$s$-wave approximation'', $F(\vb{k}) = F(k)$, where $k = \| \vb{k} \|$, because the first-order quantities explicitly depend on the anisotropic quantity $g(\vb{k})$.
%Instead, the full 2D $\vb{k}$ dependence of the THz response is treated in radial coordinates, using an angular %harmonic expansion:

We expand the fluctuation fields in angular harmonics:
\begin{equation}
	\Psi^{(1)} (\vb{k}) = \Psi^{(1)} (k,\theta_k) = \sum_{m \in \mathbb{Z}} \Psi^{(1)} (k,m) e^{im\theta_k}
%	\quad , \quad
\end{equation}
\begin{equation}
	\Psi^{(1)}(k,m) = \frac{1}{2\pi}\int_{0}^{2\pi} d\theta_k \Psi^{(1)} (k,\theta_k) e^{-im\theta_k}
	\label{eq:modeexpans}
\end{equation}
where $\Psi^{(1)}$ stands for $p^{(1)}_{e h}$, $f^{(1)}_{e}$, and $E_{\ell \lambda}^{(1)}$. (Since $E_{\ell \lambda}^{(1)}$ does not depend on $\vb{k}$, all components with $m \neq 0$ equal zero.)
%For the steady-state laser quantities, however, the $s$-wave approximation is made. Therefore the zeroth-order quantities have only the $m=0$ harmonic.
%Expansions of the form in Eq.\ \eqref{eq:modeexpans} are made for $p^{(1)}_{e h} (\vb{k},t)$, $f^{(1)}_{e} (\vb{k},t)$, and $E_{\ell \lambda}^{(1)} (t)$ in Eqs.\ \eqref{eq:delta-P-dot}--\eqref{eq:delta-E-star-dot}. These equations then contain sums over the harmonic indices $m$.
%However, it is possible to diagonalize these equations in the mode indices by applying the operator $
%These sums can be eliminated by applying the operator $\int_{0}^{2\pi} \frac{\mathrm{d} \theta_k}{2\pi} e^{-in\theta_k}$. This operation yields Coulomb matrix elements $V_{k,k^{\prime}}^{m,m^{\prime}}$, which are shown in Appx.\ \ref{appx:coulmel} to also be diagonal in the angular harmonic indices. The resulting diagonal elements are called $V_{k,k^{\prime}}^{m}$. Then the angular harmonics of the THz response are uncoupled:
% Check this:
%We also expand the THz vector potential in terms of its two orthogonal linear polarization components, $A_{Tx}$ and $A_{Ty}$, corresponding to $\bm{\epsilon}_{T} = \bm{x}$ and $=\bm{y}$, respectively. We define the THz coupling constant $g_B \equiv \frac{e\hbar}{2m_{r}c} = \frac{m_0}{m_r} \mu_B$, $\mu_B$ being the Bohr magneton.
% Rewrite the text from here on in this paragraph:
Expanding Eqs.\ \eqref{eq:delta-P-dot}--\eqref{eq:delta-E-star-dot} in the angular harmonics, we obtain the equations for each $m$ component of the fluctuation
%In the time domain, the linearized equations comprise the linearized equation for
of the interband polarization,
\begin{widetext}
\begin{eqnarray}
	i\hbar \frac{\partial }{\partial t} p^{(1)}_{e h}(k,m,t) &=&\left( \frac{%
		\hbar ^{2}k^{2}}{2m_{r}}+E_{g}
	-2
	\int_{0}^{\infty} \frac{k^{\prime} \mathrm{d}k^{\prime}}{2\pi} V_{k,k^{\prime}}^{0} f^{(0)}_{e} (k^{\prime})
	-\hbar \omega_{\ell}-i\gamma \right) p^{(1)}_{e h}(k,m,t)  \nonumber \\
	&&-2 p^{(0)}_{e h}(k) \int_{0}^{\infty} \frac{k^{\prime} \mathrm{d} k^{\prime}}{2\pi} V_{k,k^{\prime}}^{m}  f^{(1)}_{e}(k^{\prime},m,t) \nonumber \\
	&&-\left[ 1-2f^{(0)}_{e}(k)\right] \left[ \sum_{\lambda} \Gamma^{\lambda}_{e h} (k)E_{\ell \lambda}^{(1)} \delta_{0,m} +\int_{0}^{\infty} \frac{k^{\prime} \mathrm{d} k^{\prime}}{2\pi} V_{k,k^{\prime}}^{m} p^{(1)}_{e h}(k^{\prime},m,t)\right]  \nonumber
	\\
	&&+2f^{(1)}_{e}(k,m,t) \left[ \sum_{\lambda} \Gamma^{\lambda}_{e h} (k)E_{\ell \lambda}^{(0)} + \int_{0}^{\infty} \frac{k^{\prime} \mathrm{d} k^{\prime}}{2\pi} V_{k,k^{\prime}}^{0} p^{(0)}_{e h}(k^{\prime})\right] \nonumber \\
	&& + g_B k p^{(0)}_{e h}(k) \left[A_{Tx}(t) \left(\delta_{1,m} + \delta_{-1,m} \right) -i A_{Ty}(t) \left( \delta_{1,m} - \delta_{-1,m} \right) \right]   \label{eq:delta-P-dot-ang}
\end{eqnarray}%
%the linearized equation for
the complex conjugate of the interband polarization,
\begin{eqnarray}
	-i\hbar \frac{\partial }{\partial t}p^{(1) \ast}_{e h}(k,-m,t) &=&\left(
	\frac{\hbar ^{2}k^{2}}{2m_{r}}+E_{g}
	-2 \int_{0}^{\infty} \frac{k^{\prime} \mathrm{d}k^{\prime}}{2\pi} V_{k,k^{\prime}}^{0} f^{(0)}_{e} (k^{\prime})
	-\hbar \omega_{\ell}+i\gamma \right) p^{(1) \ast}_{e h}(k,-m,t)  \nonumber \\
	&&- 2 p^{(0) \ast}_{e h}(k) \int_{0}^{\infty} \frac{k^{\prime} \mathrm{d} k^{\prime}}{2\pi} V_{k,k^{\prime}}^{m} f^{(1)}_{e}(k^{\prime},m,t)   \nonumber \\
	&& -\left[ 1-2f^{(0)}_{e}(k)\right] \left[ \sum_{\lambda} \Gamma^{\lambda \ast}_{e h} (k)E_{\ell \lambda}^{(1)\ast} \delta_{m,0}
	+\int_{0}^{\infty} \frac{k^{\prime} \mathrm{d} k^{\prime}}{2\pi} V_{k,k^{\prime}}^{m} p^{(1) \ast}_{e h}(k^{\prime},-m,t)\right]  \nonumber \\
	&& +2f^{(1)}_{e}(k,m,t) \left[ \sum_{\lambda} \Gamma^{\lambda \ast}_{e h} (k)E_{\ell \lambda}^{(0)\ast}+ \int_{0}^{\infty} \frac{k^{\prime} \mathrm{d} k^{\prime}}{2\pi} V_{k,k^{\prime}}^{0} p^{(0) \ast}_{e h}(k^{\prime })\right] \nonumber \\
	&& + g_B k p^{(0) \ast}_{e h}(k) \left[A_{Tx}^{\ast}(t) \left(\delta_{1,m} + \delta_{-1,m} \right) -i A_{Ty}^{\ast}(t) \left( \delta_{1,m} - \delta_{-1,m} \right) \right] ,  \label{eq:delta-P-star-dot-ang}
\end{eqnarray}%
%the linearized equation for
the distribution function,
\begin{eqnarray}
	i\hbar \frac{\partial }{\partial t}f^{(1)}_{e}(k,m,t) &=&\left[ \sum_{\lambda} \Gamma^{\lambda \ast}_{e h} (k)E_{\ell \lambda}^{(0)\ast}+ \int_{0}^{\infty} \frac{k^{\prime} \mathrm{d} k^{\prime}}{2\pi} V_{k,k^{\prime}}^{0} p^{(0) \ast}_{e h}(k^{\prime })\right] p^{(1)}_{e h}(k,m,t)  \nonumber \\
	&&+\left[ \sum_{\lambda} \Gamma^{\lambda \ast}_{e h} (k)E_{\ell \lambda}^{(1)\ast} \delta_{m,0} + \int_{0}^{\infty} \frac{k^{\prime} \mathrm{d} k^{\prime}}{2\pi} V_{k,k^{\prime}}^{m} p^{(1) \ast}_{e h}(k^{\prime},-m,t)\right] p^{(0)}_{e h}(k)  \nonumber \\
	&&-\left[ \sum_{\lambda} \Gamma^{\lambda}_{e h} (k)E_{\ell \lambda}^{(0)}+ \int_{0}^{\infty} \frac{k^{\prime} \mathrm{d} k^{\prime}}{2\pi} V_{k,k^{\prime}}^{0} p^{(0)}_{e h}(k^{\prime})\right]
	p^{(1) \ast}_{e h}(k,-m,t)  \nonumber \\
	&&-\left[ \sum_{\lambda} \Gamma^{\lambda}_{e h} (k)E_{\ell \lambda}^{(1)} \delta_{m,0} +\int_{0}^{\infty} \frac{k^{\prime} \mathrm{d} k^{\prime}}{2\pi} V_{k,k^{\prime}}^{m} p^{(1)}_{e h}(%
	k^{\prime},m,t)\right] p^{(0) \ast}_{e h}(k)  \nonumber \\
	&&-i \gamma_{f}  f^{(1)}_{e}(k,m,t)
	\label{eq:delta-f-dot-ang}
\end{eqnarray}%
%the linearized equation for
the cavity field,
\begin{equation}
	i\hbar \frac{\partial }{\partial t}E_{\ell \lambda}^{(1)} (t) =\left( \hbar \omega _{cav}-\hbar
	\omega _{\ell}-i\gamma _{cav}\right) E_{\ell \lambda}^{(1)} (t) - N_{QW} \sum_{eh} \int_{0}^{\infty} \frac{k \mathrm{d}k}{2\pi} \Gamma^{\lambda \ast}_{e h} (k)p^{(1)}_{e h}(k,m=0,t)
	\label{eq:delta-E-dot-ang}
\end{equation}%
%and the linearized equation for
and the complex conjugate of the cavity field,
\begin{equation}
	-i\hbar \frac{\partial }{\partial t}E_{\ell \lambda}^{(1)\ast} (t) =\left( \hbar \omega_{cav}-\hbar \omega _{\ell}+i\gamma _{cav}\right) E_{\ell \lambda}^{(1)\ast} (t) - N_{QW} \sum_{eh} \int_{0}^{\infty} \frac{k \mathrm{d}k}{2\pi} \Gamma^{\lambda}_{e h} (k)p^{(1) \ast}_{e h}(k,m=0,t)   \label{eq:delta-E-star-dot-ang}
\end{equation}
\end{widetext}
In Eqs.\ \eqref{eq:delta-P-dot-ang} and \eqref{eq:delta-P-star-dot-ang}, we define the THz coupling constant $g_B \equiv \frac{e\hbar}{2m_{r}c} = \frac{m_0}{m_r} \mu_B$, $\mu_B$ being the Bohr magneton. We have chosen a linear polarization basis along the $x$ and $y$ axes for the THz field, writing the components as $A_{Tx}$ and $A_{Ty}$.
The angular momentum component of the screened Coulomb potential $V_{k,k^{\prime}}^{m}$ is given by
\begin{eqnarray}
	V_{k,k^{\prime}}^{m} &=& \frac{e^{2}}{\epsilon_{b}} \int_{0}^{2 \pi} \mathrm{d} \theta \frac{e^{- i m \theta}}{\sqrt{k^{2} + k^{\prime^{2}} - 2 k k^{\prime} \cos \theta} + \kappa_{0}} \quad , \quad \nonumber \\
	V^{c}_{| \vb{k} - \vb{k}^\prime |} &=& \sum_{m \in \mathbb{Z}} V_{k,k^{\prime}}^{m} e^{i m (\theta_k-\theta_k')}
	\label{eq:vkkpmfin}
\end{eqnarray}
where $\vb{k} = (k,\theta_k)$. The symmetry relations $V_{k,k^{\prime}}^{-m} = V_{k,k^{\prime}}^{m}$ and $V_{k,k^{\prime}}^{m} = V_{k^{\prime},k}^{m}$ are satisfied.
For an alternative formulation of Eq.\ \eqref{eq:vkkpmfin} and its evaluation in terms of elliptic integrals, see Appendix \ref{appx:coulmel}.

Eqs.\ \eqref{eq:delta-P-dot-ang}--\eqref{eq:delta-E-star-dot-ang} show explicitly that the equations for different angular momenta are decoupled. Since the THz source terms contain only the harmonics $m = \pm 1$, only $p^{(1)}_{e h}(k,m = \pm 1,t)$ and $f^{(1)}_{e}(k,m=\pm 1, t)$ are excited in the response. As shown below, the THz conductivity $\sigma_{T} (\omega)$ also contains only
$f^{(1)}_{e} (k,m = \pm 1, t)$. So the equations for $m = \pm 1$ form a closed set sufficient for the THz linear response problem.
Since there is no THz source term for the $m=0$ harmonics,
Eqs.\ \eqref{eq:delta-P-dot-ang}--\eqref{eq:delta-E-star-dot-ang} for $m=0$ are homogeneous, and we take as the solution $f^{(1)}_{e}(k,m = 0,t) = 0$, $p^{(1)}_{e h}(k,m = 0,t) = 0$, and  $E^{(1)}_{\ell \lambda} = 0$.
This is consistent with the selection rule that the cavity photon fluctuation $E^{(1)}_{\ell \lambda}$ belongs to the $m=0$ sector and so does not appear in the THz ($m = \pm 1$) response.

\subsection{Linear THz conductivity}

The reflectivity and/or transmissivity spectra of the THz probe are typically measured to study the linear resonse of the system. These spectra can be expressed in terms of the conductivity tensor that connects the THz electric field to the induced current. In this subsection, we relate the conductivity to the density response formulated above.

The THz field propagates normally to the QW, i.e.\ $\mathbf{E}_{T}(\vb{r},t)$ is a plane wave with wavevector $\vb{q} \parallel \hat{z}$ and linear polarization in the $\hat{x}$ or $\hat{y}$ directions, $\bm{\epsilon}_{\nu} = \hat{x}$ or $=\hat{y}$. It induces a two-dimensional current $\vb{J}^{(1)}$ in the quantum well. Each (linearly polarized) component of the current,
$J_{\nu}^{(1)} = \vb{J}^{(1)} \cdot \bm{\epsilon}_{\nu}$,
consists of a paramagnetic part $J_{\nu}^{p(1)}$ and a diamagnetic part $J_{\nu}^{d(1)}$, which are described in more detail in Ref.\ \onlinecite{spotnitz-etal.2021}.
In the frequency domain, the THz field and the induced current in our 2D isotropic setting are related by the conductivity $\sigma_{T\nu}(\omega)$:
\begin{equation}
	J_{\nu}^{(1)}(\omega) = \sigma_{T\nu}(\omega) E_{T\nu}(\omega) ~ ,
	\label{eq:thz-cond-def}
\end{equation}
where $E_{T\nu}(\omega)$ is the transmitted field amplitude.
Like the current density, the conductivity can be written as a sum, $\sigma_{T \nu}(\omega) = \sigma_{T \nu}^{p}(\omega) + \sigma_{T \nu}^{d}(\omega)$, of a paramagnetic term $\sigma_{T \nu}^{p} \left(\omega \right) = J_{\nu}^{p (1)} \left(\omega \right)  / E_{T\nu} \left(\omega \right)$ and a diamagnetic term $\sigma_{T \nu}^{d} \left(\omega \right) =  J_{\nu}^{d (1)} \left(\omega \right) / E_{T\nu} \left(\omega \right)$.
%Equations \eqref{pt2eq6}, \eqref{pt2eq9}, and \eqref{pt2eq11} give
The outgoing (reflected and transmitted) THz waves are given in terms of the conductivity. We quote the result here, the derivation being given in Appx.\ A of Ref.\ \cite{spotnitz-etal.2021}.
The transmission $T(\omega)$ and reflection $R(\omega)$ coefficients ($|T(\omega)|^2$ being the transmissivity and $|R(\omega)|^{2}$ being the reflectivity) and the absorptivity $A(\omega)$ are given by
\begin{IEEEeqnarray}{rCl}
	T (\omega)  &\equiv& \frac{E_{T\nu}^{(t)}(\omega)}{E_{T\nu}^{(i)}(\omega)} = \frac{1}{1 + \beta \left(\omega \right)} \label{eq:Tomega} \\
	R (\omega)  &\equiv& \frac{E_{T\nu}^{(r)}(\omega)}{E_{T\nu}^{(i)}(\omega)} =  -\frac{\beta(\omega)}{1 + \beta (\omega)}  \yesnumber \label{eq:Romega} \\
	A (\omega) &=& 1 - |T (\omega)|^2 - |R (\omega)|^2  \yesnumber \label{eq:Aomega} \\
	&=& \frac{2 \Re \beta (\omega) }{ | 1 + \beta^R (\omega ) |^2} \nonumber
\end{IEEEeqnarray}
where
\begin{equation*}
	\beta (\omega) = \frac{2 \pi}{c \sqrt{\epsilon_b}} \sigma_{T \nu} (\omega)
\end{equation*}
where the THz field amplitudes are denoted $E_{T\nu}^{(i)}(\omega)$ for the incident, $E_{T\nu}^{(r)}(\omega)$ for the reflected, and $E_{T\nu}^{(t)}(\omega)$ for the transmitted amplitudes.

The paramagnetic and diamagnetic conductivities are calculated from Eqs.\ (25) and (26) of Ref.\ \cite{spotnitz-etal.2021} as
\begin{align}
	\sigma_{T\nu}^{p} (\omega) &= -\frac{2S_{d} g_B c}{E_{T\nu}(\omega)\mathcal{A}} \sum_{\mathbf{k}} (\mathbf{k}\cdot\bm{\epsilon}_{\nu})f^{(1)}_{e}(\mathbf{k},\omega) ~,
	\yesnumber \label{eq:condthz} \\
	\sigma_{T\nu}^{\mathrm{d}} (\omega) &=  \frac{i e^{2}S_{d}}{\omega m_{r} \mathcal{A}} \sum_{\mathbf{k}} f_e^{(0)} (\mathbf{k}) ~. \label{eq:Drudecond}
\end{align}
$S_{d}$ is the spin degeneracy factor of the conduction electrons and valence holes. In terms of the angular momentum components of $f_e^{(1)}$, $\sigma_{T\nu}^{p} (\omega)$ is written for $\nu = y$ or $=x$ as
\begin{equation}
	\sigma_{T\nu}^{p} (\omega) = -\frac{S_{d} g_B c}{E_{T\nu}(\omega)} \int_{0}^{\infty} \frac{k^{2}\mathrm{d}k}{\pi}  f^{(1)}_{e}(k,m=1,\omega) e^{i \frac{\pi}{2} \delta_{\nu,y}} ~,
	\nonumber \label{eq:condthzypol}
\end{equation}
Taking into account the different source terms for $\nu = x$ and $\nu =y$, the symmetry $\sigma_{Tx}^{p} = \sigma_{Ty}^{p}$ is obtained.
The conductivity does not depend on the phase of $\mathbf{E}_{T}$, and has the symmetry $\sigma_{T \nu}(\omega)=\sigma_{T \nu}^{\ast}(-\omega)$, which leads to $A(\omega) = A(-\omega)$ and likewise for $|R|^2$ and $|T|^2$.

The Drude model of conductivity is commonly used in phenomenological analysis of data. In this model, the entire conductivity is represented as
\begin{equation}
	\sigma_{T\nu}^{\mathrm{Drude}} (\omega) =  \frac{i e^{2} n}{(\omega + i \gamma_{D}) m_{r}} \quad , \quad n = \frac{S_{d}}{\mathcal{A}} \sum_{\mathbf{k}} f^{(0)} (\mathbf{k}) ~. \label{eq:Drudecondwgm}
\end{equation}
where $\gamma_D$ accounts for all loss and relaxation processes. The diamagnetic conductivity in Eq.\ \eqref{eq:Drudecond} agrees with $\sigma_{T\nu}^{\mathrm{Drude}} (\omega)$ except for the absence of $\gamma_D$. We think that if a loss rate is phenomenologically inserted into the diamagnetic conductivity, its interpretation should be different from that of the Drude $\gamma_D$. Since in our formulation, many or all dissipative and relaxation processes are already included in the paramagnetic conductivity, only the remaining omitted processes should be represented in a loss rate in the diamagnetic conductivity. In this paper, we assume all losses are accounted for in $\sigma_{T\nu}^{p} (\omega)$.

Eqs.\ \eqref{eq:delta-P-dot-ang}--\eqref{eq:delta-f-dot-ang} and \eqref{eq:condthz} enable an interpretation of the coherent frequency mixing which leads to the paramagnetic THz response.
The rotating wave approximation is not made for the THz frequencies, so both positive and negative frequency components $\pm\omega_{T}$ contribute to the response.
In the THz source term in Eq.\ \eqref{eq:delta-P-dot-ang} (Eq.\ \eqref{eq:delta-P-star-dot-ang}), the positive and negative THz frequency components $\omega_{T}$ and $-\omega_{T}$ both add coherently to the optical-frequency polarization $p_{eh}^{(0)}$ ($p_{eh}^{(0)\ast}$) with lasing frequency $\omega_{\ell}$ ($-\omega_{\ell}$), to give the THz-induced polarization $p_{eh}^{(1)}$ ($p_{eh}^{(1)\ast}$) with frequency $\pm \omega_{T} + \omega_{\ell}$ ($\pm \omega_{T}-\omega_{\ell}$).
%Then, in Eq.\ \eqref{eq:delta-f-dot-ang}, $p_{eh}^{(1)}$ coherently mixes frequencies with $p_{eh}^{(0)\ast}$ and $E_{\ell \lambda}^{(0)\ast}$ which have frequency $-\omega_{\ell}$ to give the $f^{(1)}$ frequency component $\omega_{T} + \omega_{\ell} - \omega_{\ell} = \omega_{T}$.
Then, in Eq.\ \eqref{eq:delta-f-dot-ang}, $p_{eh}^{(1)}$ ($p_{eh}^{(1)\ast}$) coherently mixes frequencies with $p_{eh}^{(0)\ast}$ and $E_{\ell \lambda}^{(0)\ast}$ which have frequency $-\omega_{\ell}$ ($p_{eh}^{(0)}$ and $E_{\ell \lambda}^{(0)}$ which have frequency $\omega_{\ell}$) to give the $f^{(1)}$ frequency components $\pm \omega_{T} + \omega_{\ell} - \omega_{\ell} = \pm \omega_{T}$ ($\pm \omega_{T} - \omega_{\ell} + \omega_{\ell} = \pm\omega_{T}$).
Finally, Eq.\ \eqref{eq:condthz} shows that $f_{e}^{(1)}(\omega_{T})$ gives the measurable response via the conductivity $\sigma_{T\nu}^{p}(\omega_{T})$ at the same frequency.
Thus, taking the lasing state as a given, THz transitions occur as a coherent frequency mixing process between the THz probe frequency $\omega_{T}$ and the lasing frequency $\omega_{\ell}$.
The transition process for the linear THz interaction is $\pm \omega_{T} + \omega_{\ell} - \omega_{\ell} = \pm \omega_{T}$, for the positive or negative frequency component $\pm \omega_{T}$ of the THz probe.

	As the THz interaction process includes interband energies, the positive (negative) frequency response can be identified with upper hole U$_h$ to/from upper electron U$_e$ transitions (lower hole L$_h$ to/from lower electron L$_e$ transitions), although the observable THz response, i.e., the conductivity, only occurs at intraband energies.
	Thus the resonant transitions are shown as interband in Fig.\ \ref{fig:bandstruct}.
	In sum, the positive (negative) frequency probe is resonant with the UU (LL) transitions.
	The decay continuum of eigenvalues results from UL and LU transitions.
To linear order, the intraband probe excites THz-frequency density fluctuations $f^{(1)}$ and optical frequency polarization fluctuations $p^{(1)}$, where the difference of the polarization fluctuation frequency from the lasing frequency $\omega_{\ell}$ is also in the THz. The THz probe stimulates absorption or emission at the same THz frequency, but does not change the optical light field.

\section{Mode decomposition of the response function}
% \item Solution in terms of source and mode matrices.
\label{sec:modedecomp}

Eqs.\ \eqref{eq:delta-P-dot-ang} - \eqref{eq:delta-E-star-dot-ang} are a system of linear differential equations which is diagonal in $m$, but not in $k$.
For each $m$, these equations are numerically solved on a discretized grid in $k$ space.
Since we consider only the $m = \pm 1$ channels in this paper, $E^{(1)}_{\ell \lambda} = 0$ as shown above, and Eqs.\ \eqref{eq:delta-E-dot-ang} and \eqref{eq:delta-E-star-dot-ang} can be omitted. The discretized Eqs.\ \eqref{eq:delta-P-dot-ang} - \eqref{eq:delta-f-dot-ang} can be written in the following matrix form:
%That is, the different $k$ states are coupled by the Coulomb integrals. In deriving these equations, the continuous $k$ limit was taken, but numerically, the $k$ are discretized. Then, the Coulomb integrals are equivalent to sums over $k$. Hence, Eqs.\ \eqref{eq:delta-P-dot-ang}, \eqref{eq:delta-P-star-dot-ang}, \& \eqref{eq:delta-f-dot-ang} can be written as a matrix ODE:
\begin{equation}
	i\hbar \frac{\partial }{\partial t}\vec{x}=M\vec{x}+\vec{s}(t)
	\label{eq:delta-x-dot}
\end{equation}
where $\vec{x}$ denotes the column vector
\begin{equation}
	\vec{x}(m,t)=\left(
	\begin{matrix}
		\vec{p}^{\, (1)}_{e h}(m,t) \\
		\vec{p}^{\, (1)\ast}_{e h}(-m,t) \\
		\vec{f}^{\, (1)}_{e}(m,t)
	\end{matrix}
	\right)  \label{eq:definition-x}
\end{equation}
Here $\vec{p}^{\, (1)}_{e h} (m,t)$ and $\vec{f}^{\, (1)}_{e} (m,t)$ stand for column vectors whose elements are the values of the functions at the $k$ grid points: $p^{\, (1)}_{e h} (k_i, m, t)$ and $f^{\, (1)}_{e} (k_i, m, t)$, $i = 1, \cdots, N_k$, where $N_k$ is the number of $k$ points. (We use an arrow over a variable to denote a column vector of variable values over the set of discretized $k$ points.) We write the source vector in the structural form
\begin{equation}
	\vec{s}(t)=\left(\begin{matrix}{\vec{s}}_p(t)\\-{\vec{s}}_p^{\, \ast}(t)\\{\vec{s}}_f(t)\\\end{matrix}
	\right) \ .
	\label{eq:Mblockform}
\end{equation}
$\vec{s}(t)$ contains the $A_{T}(t)$ terms in Eqs.\ \eqref{eq:delta-P-dot-ang} and \eqref{eq:delta-P-star-dot-ang}.
The dimension of the vectors $\vec{x} (t)$ and $\vec{s} (t)$ is $3 N_k \equiv N$.
The matrix $M$ is a complex-valued $N \times N$ non-Hermitian matrix that is a linear function of the steady-state solution $p^{(0)}_{e h}(k)$, $f^{(0)}_e (k)$ and $E_{\ell \lambda}^{(0)}$.
$M$ depends only on the absolute value of the angular harmonic $m$, $M = M(|m|)$.
The vector $\vec{s} (t)$ is proportional to $A_{T \nu}(t)$.
	Some symmetries of $M$ and $\vec{s}$ are given in Appx.\ \ref{appx:vector-symmetries}.

\subsection{Response function constructed from eigenvectors}

%We have a linear response problem of the type
%\begin{equation}
%i\hbar\frac{\partial\vec{x}}{\partial t}=M\vec{x}+\vec{s}(t)
%\label{eq:delta-x-dot}
%\end{equation}
%where $\vec{x}$ is the solution (column) vector in an $N$-dimensional vector space, $M$ is an $N\times\ N$ matrix, and $\vec{s}$ is a source vector.
We formulate the response function associated with Eq.\ \eqref{eq:delta-x-dot} as the inverse operator of $i\hbar\frac{\partial}{\partial t} - M$.
Assume $\vec{s}(t)$ is a pulse in time and $\vec{x}=0$ initially. Fourier transforming to frequency space gives
\begin{multline}
	\hbar\omega\vec{x}(\omega)=M\vec{x}(\omega)+\vec{s}(\omega) \quad \\
	 \Rightarrow\quad\vec{x}(\omega)=\left[\hbar\omega-M\right]^{-1}\vec{s}(\omega)\equiv F(\omega)\vec{s}(\omega)
\end{multline}
where $F(\omega)$ is the linear response matrix. Denote the eigenvalues and eigenvectors of $M$ by $\lambda_n$ and $\vec{y}_n$ respectively:
\begin{equation}
	M{\vec{y}}_n=\lambda_n{\vec{y}}_n\quad,\quad\ n=1,2,\ldots,N \label{eq:righteig}
\end{equation}
Since $M$ is not Hermitian, the set of eigenvectors may fail to span the $N$-dimensional space of $\vec{x} (t)$, making $M$ non-diagonalizable, for some values of the parameters.
But this failure of diagonalizability typically occurs only at a zero-measure set of points in parameter space.\cite{kato.95}
Called exceptional points (EP), these points are where two or more eigenvalues and eigenvectors become the same.\cite{kato.95,heiss.04,hanai-etal.19,hanai-littlewood.20}
In our computations, we do not set the parameters at exactly an EP but infer the presence and location of an EP by the behavior of the eigenvalues and eigenvectors nearby. We therefore proceed with our formulation assuming the parameters are
outside of the EP set.
Construct the matrix $U$ from the eigenvectors as column vectors arranged side by side:
\begin{equation}
	U=\left(\begin{matrix}{\vec{y}}_1{\vec{y}}_2\cdots{\vec{y}}_N\end{matrix}\right)
	\label{eq:Ueigvecrep}
\end{equation}
Since the eigenvectors are linearly independent, $U$ is invertible, and $M$ is diagonalized as
\begin{equation}
	M=UDU^{-1}\quad,\quad\ D= \begin{pmatrix}\lambda_1&0&\cdots&0\\0&\lambda_2&\cdots&0\\\vdots&\vdots& \ddots &\vdots \\ 0&0&\cdots&\lambda_N \\ \end{pmatrix}
	\label{eq:Mdiag}
\end{equation}
The linear response matrix can be written as
\begin{equation}
	F(\omega)=\left[\hbar\omega-M\right]^{-1}=U\left[\hbar\omega-D\right]^{-1}U^{-1}
	\label{eq:freq-resp-matrix}
\end{equation}
In component form, it is
\begin{equation}
	F_{ij}(\omega)=\sum_{n=1}^{N}\frac{U_{in}(U^{-1})_{nj}}{\hbar\omega-\lambda_n}
	\label{eq:freq-resp-elem}
\end{equation}
(If the eigenvalue $\lambda_n$ of a mode is real-valued, the denominator should be $\hbar\omega-\lambda_n+i\eta$, $\eta \downarrow 0$.)

Returning to the time domain, Eq.\ \eqref{eq:delta-x-dot} is similarly expanded in the eigenvectors as
\begin{equation}
	i\hbar\frac{\partial\vec{x}}{\partial t}=UDU^{-1}\vec{x}+\vec{s}(t)
\end{equation}
Multiplying from the left by $U^{-1}$ gives
\begin{equation}
	i\hbar\frac{\partial\vec{b}}{\partial t}=D\vec{b}+\vec{T}(t)\quad,\quad\vec{b}=U^{-1}\vec{x}\ ,\ \vec{T}=U^{-1}\vec{s}
	\label{eq:diag-vector-time}
\end{equation}
Component-wise, Eq.\ \eqref{eq:diag-vector-time} is
\begin{equation}
	i\hbar\frac{\partial b_n}{\partial t}=\lambda_nb_n+T_n(t)\quad,\quad\ n=1,\cdots,N
	\label{eq:diag-elem-time}
\end{equation}
If at the initial time $t_0$, $x(t_0)=0$, and the source pulse comes afterwards, then the solution to Eq.\ \eqref{eq:diag-elem-time} is
\begin{equation}
	b_n(t)=-\frac{i}{\hbar}\int_{t_0}^{\infty}{dt^{\prime}\ \theta(t-t^{\prime})e^{-i\lambda_n(t-t^{\prime})/\hbar}T_n(t^{\prime})}
\end{equation}
or, in matrix form
\begin{IEEEeqnarray*}{rCl}
	\vec{b}(t) &=& -\frac{i}{\hbar}\int_{t_0}^{\infty}{dt^{\prime}\ \theta(t-t^{\prime})C(t-t^{\prime})\vec{T}(t^{\prime})}\quad,\quad\ \\
	C(t-t^{\prime}) &=& \left(\begin{matrix}e^{- \frac{i}{\hbar} \lambda_1 (t-t^{\prime})}&0&\cdots&0 \\ 0&e^{- \frac{i}{\hbar} \lambda_2(t-t^{\prime})}&\cdots&0 \\ \vdots&\vdots& \ddots &\vdots\\ 0&0&\cdots&e^{- \frac{i}{\hbar} \lambda_N(t-t^{\prime})} \end{matrix}\right)
\end{IEEEeqnarray*}
This gives the solution as
\begin{equation}
	\vec{x}(t)=\int_{t_0}^{\infty}{dt^{\prime}\  F(t-t^{\prime})\vec{s}(t^{\prime})}
\end{equation}
where
\begin{IEEEeqnarray}{rCl}
	F(t-t^{\prime}) &=& -\frac{i}{\hbar}\theta(t-t^{\prime})UC(t-t^{\prime})U^{-1}\quad, \\
	F_{ij}(t-t^{\prime}) &=& -\frac{i}{\hbar}\theta(t-t^{\prime})\sum_{n=1}^{N} U_{in} e^{-\frac{i}{\hbar} \lambda_n(t-t^{\prime})}(U^{-1})_{nj} \nonumber
\end{IEEEeqnarray}
One can verify that the Fourier transform of the time-domain response function $F(t-t^{\prime})$ is the frequency-domain response function defined in Eqs.\
\eqref{eq:freq-resp-matrix} and \eqref{eq:freq-resp-elem}.

\subsubsection*{The response function expressed in terms of the left and right eigenvectors}

$U^{-1}$ can be expressed in terms of the left eigenvectors of $M$. The left eigenvectors $\vec{z}_n, n=1,\cdots,N$ are defined as
\begin{equation}
	{\vec{z}}_n^{\, \dag} M={\widetilde{\lambda}}_n{\vec{z}}_n^{\, \dag} \qquad \text{or} \qquad M^T{\vec{z}}_n^{\, \ast}={\widetilde{\lambda}}_n{\vec{z}}_n^{\, \ast}
	\label{eq:Nailefteigenvecconv}
\end{equation}
with a set of left eigenvalues ${\widetilde{\lambda}}_n$.
Some simple properties of the left eigenmodes are as follows.
%\subsubsection*{Properties of left eigenvectors and eigenvalues:}
The sets of left eigenvalues and right eigenvalues are the same because a matrix and its transpose have equal determinants: $\mathrm{det}(\lambda-M^T)=\mathrm{det}(\lambda-M)$.
A left eigenvector and a right eigenvector belonging to two different eigenvalues are orthogonal to each other. Orthogonality is defined in the usual way as in quantum mechanics: two vectors $\vec{a}$ and $\vec{b}$ are orthogonal if
\begin{equation*}
	{\vec{a}}^{\, \dag}\vec{b}=\sum_{i}{a_i^{\ast} b_i=0}.
\end{equation*}
The orthogonality proof is similar to that in quantum mechanics. If $M{\vec{y}}_n=\lambda_n{\vec{y}}_n$ and ${\vec{z}}_k^{\, \dag}M=\lambda_k{\vec{z}}_k^{\, \dag}$, then ${\vec{z}}_k^{\, \dag}M{\vec{y}}_n=\lambda_k{\vec{z}}_k^{\, \dag}{\vec{y}}_n=\lambda_n{\vec{z}}_k^{\, \dag}{\vec{y}}_n$. If $\lambda_k\neq\lambda_n$, then ${\vec{z}}_k^{\, \dag}{\vec{y}}_n=0$.
Eigenvectors belonging to a degenerate eigenvalue can be orthogonalized within the degenerate subspace.

Returning to the representation of $U^{-1}$, we normalize the eigenvectors by requiring
\begin{equation*}
	{\vec{z}}_n^{\, \dag}{\vec{y}}_n=1
\end{equation*}
for each $n$.
Then $U^{-1}$ is given by
\begin{equation} U^{-1}=\left(\begin{matrix}&{\vec{z}}_1^{\, \dag}&\\&{\vec{z}}_2^{\, \dag}&\\&\vdots&\\&{\vec{z}}_N^{\, \dag}&\\\end{matrix}\right)
\end{equation}
The orthonormalization $\vec{z}_{k}^{\, \dag} \vec{y}_{n} = \delta_{kn},$ enforces $U^{-1} U = U U^{-1} = I$.
The linear response function can be written as
\begin{equation}
	F(\omega)=\sum_{n=1}^{N}\frac{{\vec{y}}_n{\vec{z}}_n^{\, \dag}}{\hbar\omega-\lambda_n}\quad,\quad\ F_{ij}(\omega)=\sum_{n=1}^{N}\frac{y_{n,i}z_{n,j}^{\ast}}{\hbar\omega-\lambda_n} \ .
\end{equation}
The corresponding solution $\vec{x}(\omega)$ is
\begin{equation}
	\vec{x}(\omega)=\sum_{n=1}^{N}{c_n(\omega)}{\vec{y}}_n\quad,\quad\ c_n(\omega)=\frac{{\vec{z}}_n^{\, \dag}\vec{s}(\omega)}{\hbar\omega-\lambda_n} \ .
	\label{eq:linrespvecomeigvecexp}
	% Linear response vector in omega space: eigenvector expansion
\end{equation}
The corresponding expressions in the time domain are
\begin{IEEEeqnarray}{rCl}
		F(t-t^{\prime}) &=& -\frac{i}{\hbar}\theta(t-t^{\prime})\sum_{n=1}^{N}{{\vec{y}}_ne^{-i\lambda_n(t-t^{\prime})/\hbar}
		{\vec{z}}_n^{\, \dag}}\quad,\quad\ \\
		F_{ij}(t-t^{\prime}) &=& -\frac{i}{\hbar}\theta(t-t^{\prime})\sum_{n=1}^{N}{y_{n,i}e^{-i\lambda_n(t-t^{\prime})/\hbar}
		z_{n,j}^{\ast}} \ ,
\end{IEEEeqnarray}
and the solution is
\begin{multline}
	\vec{x}(t)=\sum_{n=1}^{N} c_n(t){\vec{y}}_n \quad, \quad \\
	 c_n(t)=-\frac{i}{\hbar}\int_{t_0}^{\infty}{\theta(t-t^{\prime})e^{-i\lambda_n(t-t^{\prime})/\hbar}
		{\vec{z}}_n^{\, \dag}\vec{s}(t^{\prime})} \ .
\end{multline}

%\section{Phase and amplitude modes. Angular momentum expansion of the amplitude-phase representation of the order %parameter in linear response}

\section{Phase and amplitude representation of the response function}
\label{sec:phasampresp}

%\subsection{Switching to amplitude-phase representation}
The response function $F(t)$ (or $F(\omega)$) is expressed in Sec.\ \ref{sec:modedecomp} as a function of the eigenvectors of the
interband polarization fluctuation $\vec{p}^{\, (1)}_{e h} (m,t)$, its complex conjugate, and the density fluctuation.
Since phase and amplitude modes of the coherent steady state are of physical interest, (a.k.a., Goldstone and Higgs modes, resp.,)
the interpretation of our numerical results will be helped by switching to a representation in terms of the phase and amplitude of $\vec{p}^{\, (1)}_{e h} (m,t)$.
This is formulated in this section.

\subsubsection*{Time domain}

Write the polarization in amplitude-phase form and write the amplitude and phase as sums of unperturbed and response terms:
\begin{multline}
	p_{eh}(\vb{k},t)=R(\vb{k},t) \ e^{i \phi(\vb{k},t)} \\
	= \left[ R^{(0)}(\vb{k},t) +R^{(1)}(\vb{k},t) \right] e^{i \left[\phi^{(0)} (\vb{k},t)+\phi^{(1)}(\vb{k},t) \right]}
\end{multline}
Linearize in the first-order response terms:
\begin{widetext}
	\begin{equation}
		p_{eh}(\vb{k},t)=\left[R^{(0)}(\vb{k},t)+R^{(1)}(\vb{k},t)+iR^{(0)}(\vb{k},t)\phi^{(1)}(\vb{k},t)\right] \ e^{i\phi^{(0)}(\vb{k},t)} + \cdots
	\end{equation}
	Expand $R^{(1)}$ and $\phi^{(1)}$ in angular momentum:
	\begin{equation}
		p_{eh}(\vb{k},t)=\left[R^{(0)}(k,t)+\sum_{m}{\left(R_m^{(1)}(k,t)+iR^{(0)}(k,t)\phi_m^{(1)}(k,t)\right)\   e^{im\theta_k}}\right]\   e^{i\phi^{(0)}(k,t)} \label{eq:R-phi-linear}
	\end{equation}
	Comparing Eq.\ \eqref{eq:R-phi-linear} with
	\begin{equation}
		p_{eh}(\vb{k},t)=p^{(0)}_{eh}(k,t) +
		\sum_{m} p^{(1)}_{eh}(k,m,t) e^{im\theta_k}
	\end{equation}
	gives ($R^{(0)}$ and $\phi^{(0)}$ being isotropic)
	\begin{equation}
		p_{e h}^{(1)}(k,m,t)=\  e^{i\phi^{(0)}(k,t)} \left( R_{m}^{(1)}(k,t)+iR^{(0)}(k,t)\phi_m^{(1)}(k,t)\right)
		\label{eq:R1-phi1-t}
	\end{equation}
\end{widetext}
Note that because $R^{(1)}(\vb{k},t)$ and $\phi^{(1)}(\vb{k},t)$ are real,
%$R_0^{(1)}(k,t)$ and $\phi_0^{(1)}(k,t)$ also real, and
\begin{IEEEeqnarray}{rCl}
	R_{-m}^{(1)}(k,t) &=& R_m^{(1)\ast}(k,t) \quad , \quad \nonumber \\
	\phi_{-m}^{(1)}(k,t) &=& \phi_m^{(1)\ast}(k,t) \label{eq:m-sym-t}
\end{IEEEeqnarray}

\subsubsection*{Frequency domain}

The unperturbed laser is assumed to be in a steady state ($e^{-i\omega_\ell t}$ has been taken out) so that $R^{(0)}$ and $\phi^{(0)}$ are time-independent.
Fourier transform the response quantities with respect to time.
%\\ $p_{eh,m}^{(1)}(k,\omega) =\int dt\, e^{i\omega t} p_{eh,m}^{(1)}(k,t)$       etc. \\
The relations Eq.\ \eqref{eq:m-sym-t} between $m$ and $-m$ components in the time domain translate to the following relations in the frequency domain:
\begin{IEEEeqnarray}{rCl}
	R_{-m}^{(1)}(k,\omega) &=& R_m^{(1)\ast}(k,-\omega)\quad,\quad \nonumber \\
	\phi_{-m}^{(1)}(k,\omega) &=& \phi_m^{(1)\ast}(k,-\omega)
	\label{eq:m-sym-omega} \ .
\end{IEEEeqnarray}
The Fourier transform of Eq.\ \eqref{eq:R1-phi1-t} is
\begin{multline}
	p_{e h}^{(1)}(k,m,\omega)=   \\
	e^{i\phi^{(0)}(k)}  \left( R_{m}^{(1)}(k,\omega)+iR^{(0)}(k)\phi_{m}^{(1)}(k,\omega) \right) \label{eq:R1-phi1-omega}
\end{multline}
From Eqs.\ \eqref{eq:m-sym-omega} and \eqref{eq:R1-phi1-omega}, we have
\begin{multline}
	p_{e h}^{(1)\ast}(k,-m,-\omega)= \\
	  e^{-i\phi^{(0)}(k)} \left( R_{m}^{(1)}(k,\omega)-iR^{(0)}(k)\phi_{m}^{(1)}(k,\omega) \right) \label{eq:R1-phi1-omega-prime}
\end{multline}
Solving for $R_m^{(1)}(k,\omega)$ and $\phi_m^{(1)}(k,\omega)$ from Eqs.\ \eqref{eq:R1-phi1-omega}
and \eqref{eq:R1-phi1-omega-prime} gives
\begin{equation}
	R_m^{(1)}(k,\omega)= \frac{1}{2} \left({\widetilde{p}}_{e h}^{(1)}(k,m,\omega)+{\widetilde{p}}_{e h}^{(1)\ast}(k,-m,-\omega)\right)
	\label{eq:rm1komofp}
\end{equation}
\begin{multline}
	\phi_m^{(1)}(k,\omega) = \\
	\frac{1}{2iR^{(0)}(k)}\left({\widetilde{p}}_{e h}^{(1)}(k,m,\omega)-{\widetilde{p}}_{e h}^{(1)\ast}(k,-m,-\omega)\right)
	\label{eq:phimkomofrp}
\end{multline}
where
\begin{equation*}
	{\widetilde{p}}_{eh}^{(1)}(k,m,\omega)=\  e^{-i\phi^{(0)}(k)} p_{eh}^{(1)} (k,m,\omega)
\end{equation*}
The amplitude response $R_m^{(1)}(k,\omega)$ and the phase response $\phi_{m}^{(1)}(k,\omega)$ can be computed from $p_{eh}^{(1)}(k,m,\omega)$ through Eqs.\ \eqref{eq:rm1komofp} and \eqref{eq:phimkomofrp}.

\begin{figure*}
	\centering
	\includegraphics{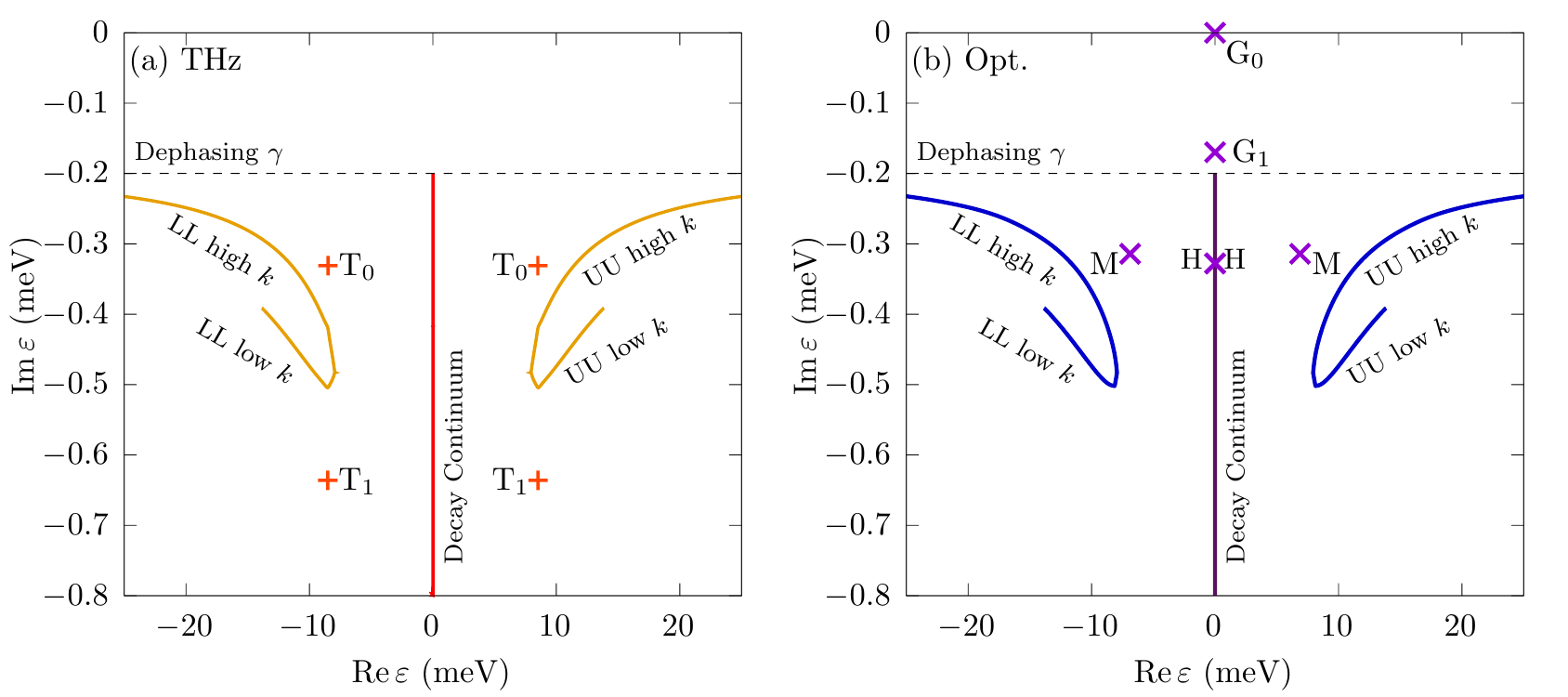}
	\caption{%
		(Color online.)
		The linear response eigenvalues $\varepsilon_{n}$
			(written as $\lambda_{n}$ in Eq.\ \eqref{eq:righteig} and Sec.\ \ref{sec:modedecomp})
		in the case of (a) an intraband THz probe and (b) an interband optical probe.
		The discrete modes differ between (a) and (b), while the spectral continua (UU, LL, and decay) are nearly identical.
		The eigenvalues are symmetric under $\Re \varepsilon \to - \Re \varepsilon$.
		The T modes are the only discrete modes predicted with an intraband probe, while an interband probe may see G, M, and H modes.
		%		The THz and optical UU and LL continua are almost identical, except for some small deviations in the vicinity of the ``hook'' at $k_{BCS}$, where low $k$ crosses into high $k$.
		%		The THz UU and LL continua differ from the optical in having sharper kinks at the $\Re \varepsilon$ of the T modes.
		(The default parameter values for all figures are given in Sec.\ \ref{sec:results}.)
	}
	\label{fig:lineigen}
\end{figure*}

\subsection{Response function}

%\section{Phase, amplitude and density modes}

The vectors $\vec{x} (m,t)$ and $\vec{s} (m,t)$ are given in block from in Eqs.\ \eqref{eq:definition-x} and \eqref{eq:Mblockform}.
Their frequency domain counterparts are
\begin{IEEEeqnarray}{rCl}
	\vec{x}(m,\omega) &=& \left(\begin{matrix}{\vec{p}}^{\,(1)}_{e h}(m,\omega)\\
		{\vec{p}}^{\, (1)\ast}_{e h}(-m,-\omega)\\
		{\vec{f}}^{\, (1)}_e (m,\omega)\\\end{matrix}\right)	%\quad,\quad\vec{\xi}(\omega)=\left(\begin{matrix}{\vec{p}}_{eh}^{(1)}(\omega)\\E^{(1)}(\omega)\\\end{matrix}\right)
		\label{eq:xomblockform}
	%and we write the source column vector as
	\quad , \quad \\
	\vec{s}(m,\omega) &=&
	\left(\begin{matrix}{\vec{s}}_{p} (m,\omega) \\
		-{\vec{s}}_p^{\, \ast}(-m,-\omega)\\
		{\vec{s}}_f(m,\omega)\\\end{matrix}\right) .
		\label{eq:somblockform}
\end{IEEEeqnarray}
Write the left and right eigenvectors as
\begin{equation}
	\vec{y}_{n} =
	\begin{pmatrix}
		\vec{X}_n \\ \vec{Y}_n \\ \vec{Z}_{n}
	\end{pmatrix}
	, \quad
	\vec{z}_{n} =
	\begin{pmatrix}
		\vec{X}_{n}^{\prime} \\ \vec{Y}_{n}^{\prime} \\ \vec{Z}_{n}^{\prime}
	\end{pmatrix},
	\label{eq:eigveccomp}
\end{equation}
respectively.
With these definitions, Eq.\ \eqref{eq:linrespvecomeigvecexp} becomes
\begin{widetext}
\begin{equation}
	\left(\begin{array}{l}
		\vec{p}^{\, (1)}_{e h}(m,\omega) \\
		\vec{p}^{\,(1) *}_{e h}(-m,-\omega) \\
		\vec{f}^{\, (1)}_e (m,\omega)
	\end{array}\right)=\sum_{n=1}^{N} \frac{1}{\hbar \omega-\lambda_{n}}\left(\begin{array}{l}
		\vec{X}_{n} \\
		\vec{Y}_{n} \\
		\vec{Z}_{n}
	\end{array}\right)\left(\vec{X}_{n}^{\prime \dagger} \vec{s}_{p}(m,\omega)-\vec{Y}_{n}^{\prime \dagger} \vec{s}_{p}^{\, *}(-m,-\omega)+\vec{Z}_{n}^{\prime \dagger} \vec{s}_{f}(m,\omega)\right)
	\label{eq:xiomofxyzsblock}
\end{equation}
%Switch to amplitude-phase basis and put in the $m$ indices:
%(include $E^{(1)}(\omega)$ in the definitions of $R_{m}^{(1)}$ \& $\phi_{m}^{(1)}$ from Eqs.\ \eqref{eq:rm1komofp} \& \eqref{eq:phimkomofrp})
%\begin{eqnarray}
%	R_m^{(1)}(k,\omega) &=& \frac{1}{2} %\left({\tilde{\xi}}_{eh,m}^{(1)}(k,\omega)+{\tilde{\xi}}_{eh,-m}^{(1)\ast}(k,-\omega)\right)
%	\label{eq:rm1komofxi}
%\\
%	\phi_m^{(1)}(k,\omega) &=& %\frac{1}{2iR^{(0)}(k)}\left({\tilde{\xi}}_{eh,m}^{(1)}(k,\omega)-{\tilde{\xi}}_{eh,-m}^{(1)\ast}(k,-\omega)\right)%
%	\label{eq:phimkomofrxi}
%\end{eqnarray}
%\begin{equation}
%\begin{array}{ll}
%	\tilde{\xi}_{m}^{(1)}(k, \omega)=e^{-i \phi^{(0)}(k)} \xi_{m}^{(1)}(k, \omega) ; & k=1, \cdots, N_{k}+1 \\
%	& \xi_{m}^{(1)}\left(N_{k}+1, \omega\right)=E^{(1)}(\omega)
%\end{array}
%\end{equation}
%$R^{(0)}\left(N_{k}+1\right), \phi^{(0)}\left(N_{k}+1\right)$ are the amplitude \& phase of $E^{(0)}$
The interband polarization is written in phase-amplitude form in
Eqs.\ \eqref{eq:rm1komofp} and \eqref{eq:phimkomofrp}.
We perform the same transformation on the source vector
\begin{eqnarray}
	\vec{s}_{R}(m,\omega)&=&\frac{1}{2}\left(\tilde{\vec{s}}_{p}(m,\omega)+\tilde{\vec{s}}_{p}^{\, *}(-m,-\omega) \right) ; \quad \tilde{s}_{p}(k,m,\omega)=e^{-i \phi^{(0)}(k)} s_{p}(k,m,\omega)
	\label{eq:srofsxitilde} \\
	\vec{s}_{\phi}(m,\omega) &=& \frac{1}{2i} \left(\tilde{\vec{s}}_{p}(m,\omega)-\tilde{\vec{s}}_{p}^{\, *}(-m,-\omega)\right)
	\label{eq:sphiofsxitilde} \\
	\Rightarrow \ \tilde{\vec{s}}_{p} (m,\omega) &=& \vec{s}_{R}(m,\omega) + i \vec{s}_{\phi}(m,\omega), \quad
	\tilde{\vec{s}}_{p}^{\, *}(-m,-\omega) = \vec{s}_{R}(m,\omega) - i \vec{s}_{\phi}(m,\omega) \ .
	\label{eq:sxitildeofsrphi}
\end{eqnarray}
For the case where $E_{\ell\lambda}^{(1)} \neq 0$, such as with an optical (interband) probe, $E_{\ell\lambda}^{(1)}$ can be written in $R$ and $\phi$ components in the same way as  $\vec{p}_{eh}^{\, (1)}$.
With Eqs.\ \eqref{eq:srofsxitilde}--\eqref{eq:sxitildeofsrphi}, the coefficient in front of $\vec{y}_n$ in Eq.\ \eqref{eq:xiomofxyzsblock} can be written as
\begin{IEEEeqnarray*}{rCl}
	c_{n} (\omega) &=& (\hbar\omega - \lambda_{n})^{-1} \left[ \vec{X}_{n}^{\prime \dagger} \vec{s}_{p}(m,\omega)-\vec{Y}_{n}^{\prime \dagger} \vec{s}_{p}^{\, *}(-m,-\omega)+\vec{Z}_{n}^{\prime \dagger} \vec{s}_{f}(m,\omega) \right] \\
	&=& (\hbar\omega - \lambda_{n})^{-1} \left[ \left(\tilde{\vec{X}}_{n}^{\prime \dagger} -\tilde{\vec{Y}}_{n}^{\prime \dagger} \right) \vec{s}_{R}(m,\omega)
	+ i \left(\tilde{\vec{X}}_{n}^{\prime \dagger} +\tilde{\vec{Y}}_{n}^{\prime \dagger} \right) \vec{s}_{\phi}(m,\omega)
	+\vec{Z}_{n}^{\prime \dagger} \vec{s}_{f}(m,\omega) \right] \yesnumber
\end{IEEEeqnarray*}
\end{widetext}
where we have defined
\begin{equation}
	\tilde{X}_{n k}^{\prime} = X_{n k}^{\prime} e^{-i \vec{\phi}^{(0)}(k) }, \quad
	\tilde{Y}_{n k}^{\prime} = Y_{n k}^{\prime} e^{i \vec{\phi}^{(0)}(k) }
	%, \quad
	%	\vec{\phi}^{(0)} = \begin{pmatrix}
		%		\phi_{P}^{(0)} (k), \phi_{E}^{(0)}
		%	\end{pmatrix}
\end{equation}
Taking the sum and difference of the first two equations in Eq.\ \eqref{eq:xiomofxyzsblock} gives
\begin{equation}
	\left(\begin{array}{l}
		\vec{R}_{m}^{(1)}(\omega) \\
		\vec{\alpha}_{m}^{\, (1)}(\omega) \\
		\vec{f}^{\, (1)}_e (m,\omega)
	\end{array}\right)=\sum_{n=1}^{N} c_{n}(\omega)\left(\begin{array}{c}
		\frac{1}{2}\left(\tilde{\vec{X}}_{n}+\tilde{\vec{Y}}_{n}\right) \\
		\frac{1}{2 i}\left(\tilde{\vec{X}}_{n}-\tilde{\vec{Y}}_{n}\right) \\
		\vec{Z}_{n}
	\end{array}\right)
	\label{eq:rphivecexp}
\end{equation}
where we have defined
\begin{IEEEeqnarray*}{rCl}
	\alpha_{m}^{\, (1)} (k,\omega) &=& R^{(0)}(k) \phi_{m}^{(1)} (k,\omega), \\
	\tilde{X}_{nk} &=& {X}_{nk} e^{-i {\phi}^{(0)} (k)}, \\
	\tilde{Y}_{nk} &=& Y_{nk} e^{i \phi^{(0)} (k)}
	.
\end{IEEEeqnarray*}
Eq.\ \eqref{eq:rphivecexp} can also be written in a response function form:
\begin{equation}
	\left(\begin{array}{l}
		\vec{R}_{m}^{(1)}(\omega) \\
		i \vec{\alpha}_{m}^{\, (1)}(\omega) \\
		\vec{f}^{\, (1)}_e (m,\omega)
	\end{array}\right)
	= \begin{pmatrix}
		F_{RR} & F_{R\phi} & F_{R f} \\
		F_{\phi R} & F_{\phi \phi} & F_{\phi f} \\
		F_{f R} & F_{f \phi} & F_{ff}
	\end{pmatrix}
	\begin{pmatrix}
		\vec{s}_{R} (m,\omega) \\
		i \vec{s}_{\phi} (m,\omega) \\
		\vec{s}_{f} (m,\omega)
	\end{pmatrix}
	\label{eq:rphivecrespfunc}
\end{equation}
where the block sub-matrices are given by
\begin{equation}
	F_{i j}(\omega)=\sum_{n=1}^{N} \frac{\vec{a}_{n i}  \vec{b}_{n j}^{\, \dagger}}{\hbar \omega-\lambda_{n}} \quad i, j=R, \phi, f \label{eq:Fijom-phas}
\end{equation}
\begin{IEEEeqnarray}{rCl}
	\left(\begin{array}{c}
		\vec{a}_{n R} \\
		\vec{a}_{n \phi} \\
		\vec{a}_{n f}
	\end{array}\right) &=& \left(\begin{array}{c}
		\frac{1}{2}\left(\tilde{\vec{X}}_{n}+\tilde{\vec{Y}}_{n}\right) \\
		\frac{1}{2}\left(\tilde{\vec{X}}_{n}-\tilde{\vec{Y}}_{n}\right) \\
		\vec{Z}_{n}
	\end{array}\right) \quad, \\
	\left(\begin{array}{c}
		\vec{b}_{n R} \\
		\vec{b}_{n \phi} \\
		\vec{b}_{n f}
	\end{array}\right) &=& \left(\begin{array}{c}
		\tilde{\vec{X}}_{n}^{\prime}-\tilde{\vec{Y}}_{n}^{\prime} \\
		\tilde{\vec{X}}_{n}^{\prime}+\tilde{\vec{Y}}_{n}^{\prime} \\
		\vec{Z}_{n}^{\prime}
	\end{array}\right) \quad .
\end{IEEEeqnarray}
%
% Note the convention is different for $\vec{a}$ in the code
%(Recall that $\vec{a}_{n R},
%\vec{a}_{n \phi}, \vec{b}_{n R},
%\vec{b}_{n \phi}$ are $(N_k +1)$-dimensional vectors and $\vec{a}_{n f}, \vec{b}_{n f}$ are $N_k$-dimensional %vectors.)
%\subsection{Phase, amplitude, and density modes}
The classification of a single mode is determined by the corresponding summand in Eq.\ \eqref{eq:Fijom-phas}.
We call a mode $n$ a pure $\begin{pmatrix}
	\text{amplitude} \\ \text{phase} \\ \text{density}
\end{pmatrix}$ mode if its
$\begin{pmatrix}
	\vec{a}_{n R} \\
	\vec{a}_{n \phi} \\
	\vec{a}_{n f}
\end{pmatrix}$
component is predominant.
If none of the $\vec{a}_{ni}$ components can be said to clearly dominate, then it is a mixed mode.
Collective modes have eigenvectors distributed over a relatively wide range of $k$ values, and eigenvalues which are discrete, rather than part of a continua.
To be a Higgs mode, a mode must be a collective amplitude mode, with $\vec{a}_{R} \gg \vec{a}_{\phi},\vec{a}_{f}$; and to be a Goldstone mode, a mode must be a collective phase mode, with $\vec{a}_{\phi} \gg \vec{a}_{R}, \vec{a}_{f}$.
In Figs.\ \ref{fig:ampl_phas_box_log}
and \ref{fig:ampl_phas_lograt}, the symbols $R$, $\alpha$, and $f$, for a given mode, with or without the superscript $^{(1)}$, refer to $\vec{a}_{R}$, $\vec{a}_{\phi}$, and $\vec{a}_{f}$ for a right eigenvector, and $\vec{b}_{R}$, $\vec{b}_{\phi}$, and $\vec{b}_{f}$ for a left eigenvector, respectively.

\section{Results and Discussion}
\label{sec:results}

In the following, we present numerical results.
Unless otherwise noted, we use the following parameter values:
number of quantum wells $N_{QW}=1$,
effective Bohr radius $a_{B}=14$ nm,
background dielectric constant $\epsilon_{b}=16.1$,
unrenormalized band gap $E_{G}=1.562$ eV;
$E_{B}^{2D}=12.8$ meV, where the exciton binding energy in 2D
is related to the one in 3D via $E_{B}^{2D}=4E_{B}^{3D}$ and $E_{B}^{3D}=\frac{\hbar ^{2}}{2m_{r}a_{B}^{2}}$ determines the reduced e-h mass,
$m_{r}^{-1}=m_{e}^{-1}+m_{h}^{-1}$ ($m_{e}=m_{h}$ in our approximation);
therefore the effective electron mass is $m_{e} = 0.121m_{0}$, where $m_{0}$ is the free electron mass;
cavity resonance frequency $\hbar \omega _{cav}=1.550$ eV,
screening wavenumber $\kappa_{0} = 9 \times 10^{-3} a_{B}^{-1}$,
dephasing $\gamma = 0.2$ meV, Fermi relaxation rate $\gamma_{F} = 2\gamma$,
pump relaxation rate $\gamma_{p}=0.4$ meV, non-radiative decay rate $\gamma_{nr}=10^{-4}$ meV, total distribution relaxation rate $\gamma_{f}= \gamma_{F} + \gamma_{p} +\gamma_{nr}= 0.8001$ meV, cavity decay rate $\gamma _{cav}=0.1$ meV,
effective electron temperature $T = 50$ K,
%$r_{\mathrm{cv}}=0.5$ nm where $d_{cv}=-er_{\mathrm{cv}}$
and interband coupling constant $\Gamma_{eh}^{\lambda} = 64.04$ peV$\cdot$m.
The default pump density is $n_{p} = 1 a_{B}^{-2}$.
In figures \ref{fig:sm1261_cond_re_dens}--\ref{fig:bcs_gap_dens},
%, \ref{fig:cond_im_r_np}, \ref{fig:cond_re_at_np}, and \ref{fig:bcs_gap_dens},
the dephasing is $\gamma = 0.5$ meV, so $\gamma_{F}= 1$ meV and $\gamma_{f}= 1.4001$ meV.

\subsection{Fluctuation Spectrum}
\label{subsec:fluct-spect}

In Fig. \ref{fig:lineigen} we show the complex eigenvalues, comparing the
 new case of intraband fluctuations (fluctuations triggered by THz fields)
 with the known spectrum \cite{binder-kwong.2021}
 for the case of interband fluctuations (i.e., fluctuations modes triggered by interband fields).
%%%% All interband modes have angular momentum $m=0$, and all intraband modes have $m=\pm 1$ (units of $\hbar$).
 The interband modes contain the Goldstone modes $G_0$, Goldstone companion modes $G_1$, discrete collective modes ($M$, $H$), a decay continuum (vertical continuum in the figure, at the lasing frequency), and positive and negative frequency spectral continua (almost horizontal in the figure) with the `hook' feature.
 See Ref.\ \onlinecite{binder-kwong.2021} for more details on the collective modes in Fig.\ \ref{fig:lineigen}(b) and on the evolution of these optically-induced fluctuation modes from below to above the lasing threshold.
   The intraband modes do not contain the Goldstone modes (because of their different angular momentum), but they do contain collective modes T$_i$ and continua similar to the interband modes.

\begin{figure*}
	% Eigenvalue vs k.
	\centering
	\includegraphics{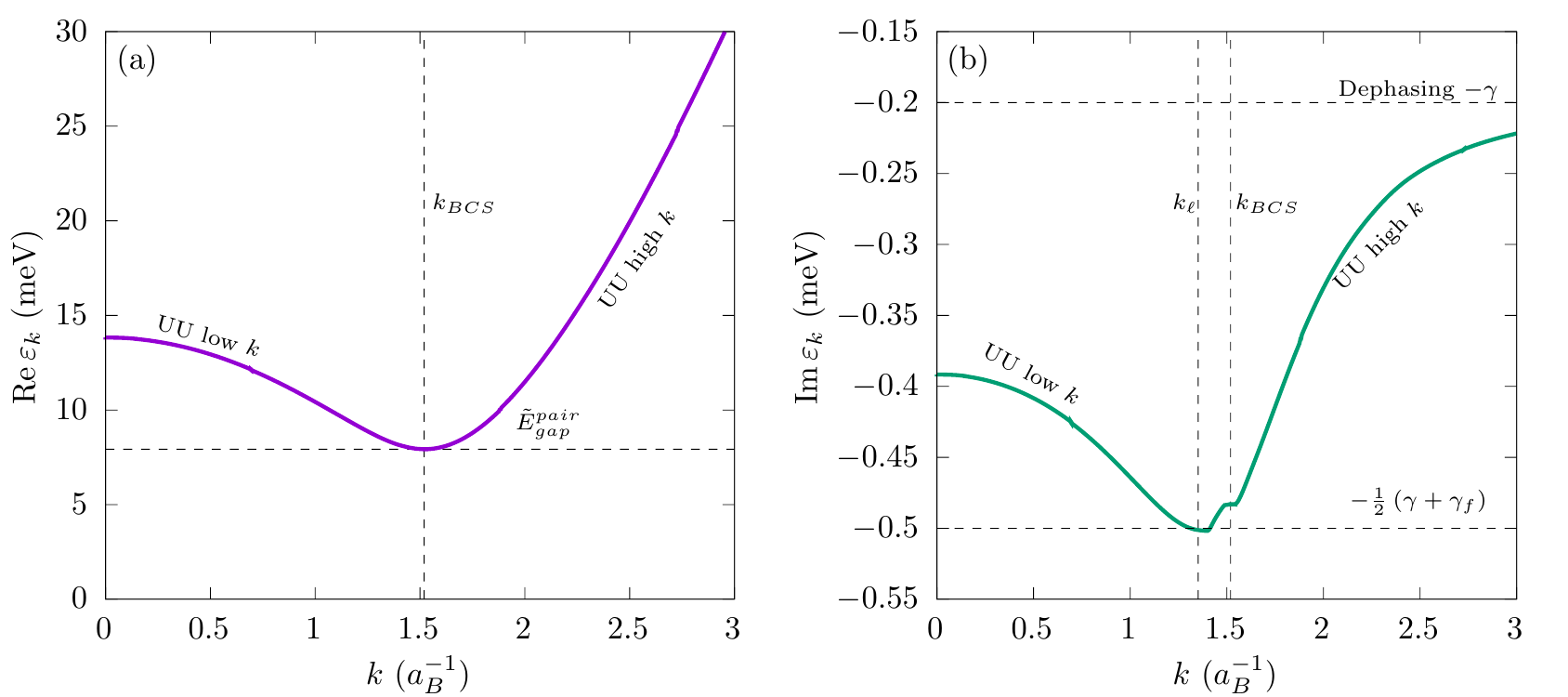}
	\caption{The real and imaginary parts of the UU continuum eigenvalues $\varepsilon_{k}$ as a function of the wavenumber $k$ of the corresponding eigenvector's magnitude peak, $\varepsilon_{k} \equiv \lambda_{n}$ s.t.\ $k = \max_{k'} |\vec{X}_{n} (k')|$, taken over all $n$ in the UU continuum. (Here, $\lambda_{n}$ and $\vec{X}_{n}$ are defined in Eqs.\ \eqref{eq:righteig} and \eqref{eq:eigveccomp}, resp.)
		$\varepsilon_{k}$ for the THz response matrix is nearly identical.
		(a) The real part is $\Re \varepsilon_{k} = 2 \tilde{E}_{k}$, twice the excitation/branch energy.
		% Note that this result would likely be modified, to $\xi_{eh}^{-} + \tilde{E}$, if we take $m_{e} \neq \m_{h}$.
		The minimum branch energy, i.e., the BCS gap $\tilde{E}_{gap}^{pair}$, occurs at $k_{BCS}$. $k_{BCS}$ divides the UU continuum into high and low $k$ regions, which are also indicated in
		\ref{fig:lineigen} and \ref{fig:bandstruct}.
		(b) $\Im \varepsilon_{k}$ is bounded above by the dephasing $-\gamma$ and below by $\approx -\frac{1}{2}\left(\gamma+\gamma_{f}\right)$, with this minimum at $\approx k_{\ell}$. A second kink occurs near $k_{BCS}$.
		In the case where $\gamma_{f}$ is taken to be $\gamma$,
		$\Im \varepsilon_{k} = -\gamma$ uniformly.}
	\label{fig:eigenergkspect}
\end{figure*}

   The UU and LL continua can be traced back to the single-particle spectrum (electronic band structure) by noting that their eigenfunctions are sharply peaked at a given wave number $k$, which is then associated with the corresponding eigenvalue $\varepsilon$.
   This gives the real part of the eigenenergies vs $k$ as $\Re \varepsilon_{k} = 2 \tilde{E}_{k}$.
   The resulting plot, figure \ref{fig:eigenergkspect}, shows the wave vector dependence of the energies of the continuum states.
   Fig.\ \ref{fig:eigenergkspect}(a) shows the real part of the energies vs $k$, in analogy to the peak positions of the continuum states in the first-order density response function shown in Fig.\ \ref{fig:sm1233_f1_abs_clrmap_no0}. The corresponding imaginary part is shown in Fig.\ \ref{fig:eigenergkspect}(b).
   It is then possible to identify the ``low $k$'' hook feature with the single-particle states created by the light-induced bands at $k<k_{BCS}$, where $k_{BCS}$ is the location of the light-induced gap in the single-particle spectrum, Fig. \ref{fig:bandstruct}.

\begin{figure}
	% T mode evolution figure.
	\centering
	\includegraphics{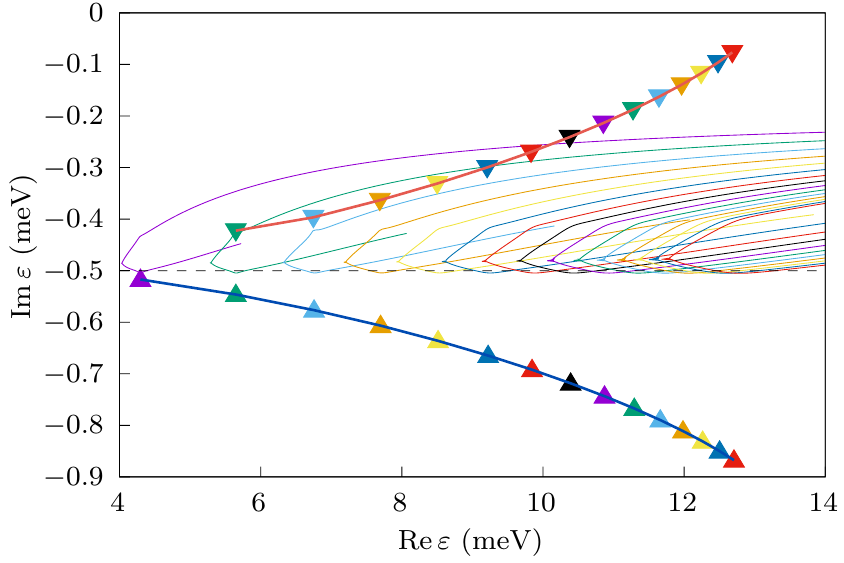}
	\caption{Plot of the eigenvalues, $\Im \varepsilon$ vs $\Re \varepsilon$, for the THz $M(m=1)$ matrix, showing the evolution of the modes T$_{0}$ (above) and T$_{1}$ (below) for color-coded pump densities $n_{p}$ from $0.6/a_B^2$ through $2/a_B^2$ in increments of $0.1/a_B^2$. The T$_{0}$ and T$_{1}$ mode energies are roughly symmetric with respect to the dashed line at $-\frac{1}{2}(\gamma+\gamma_{f})$.
		The collective (discrete) T modes are shown as triangles, connected by thick lines;
		and the UU continua are shown as thin solid lines.}
	\label{fig:thz_eigval_tmode_track}
\end{figure}

Figure \ref{fig:thz_eigval_tmode_track}, similar in format to Fig.\ \ref{fig:lineigen}(a), shows the eigenvalue and T mode evolution with increasing $n_{p}$.
	For high enough $n_{p}$, the T$_0$ modes can become unstable ($\Im \varepsilon > 0$).

\begin{figure*}
	\centering
	\includegraphics{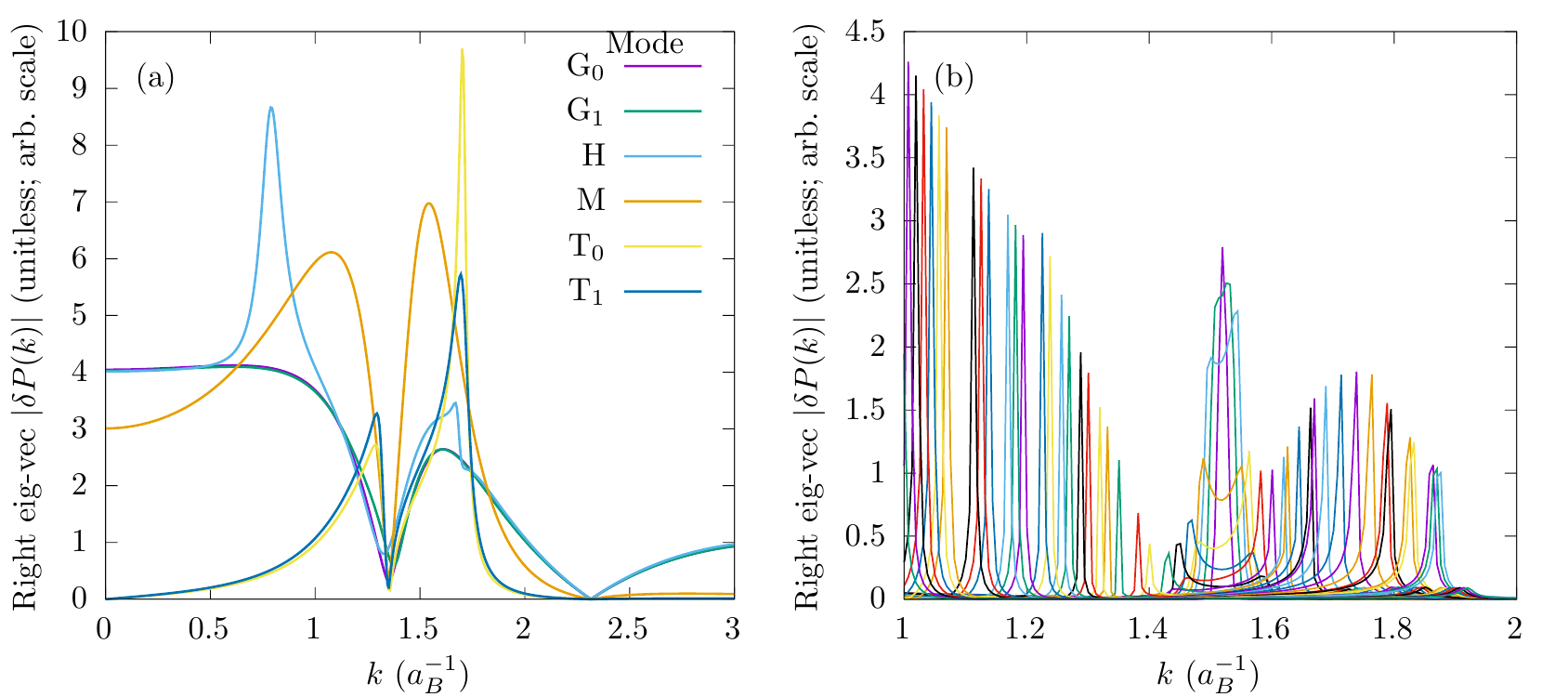}
	\caption{Plots of the magnitude of the
		%$\delta P(k)$
		$\vec{p}_{eh}^{\, (1)}$
		component of some right eigenvectors, i.e., $\vec{X}$ in the notation of Sec.\ \ref{sec:phasampresp}, and denoted $\delta P(k)$ here. The eigenvectors are unitless, and each has its own arbitrary scaling factor. Only the T modes are taken from the THz response matrix; the rest are taken from the optical probe response matrix. (See the supplement to Ref.\ \cite{binder-kwong.2021} for details of the optical probe response. The interband-probe $M$ matrix differs only in having a different constant factor on $V_{k,k'}^{0}$, in having only $m=0$ components, and therefore in having a nonzero $E_{\ell\lambda}^{(1)}$.)
		(a) Collective, discrete-eigenvalue modes. All have minima near the laser $k_{\ell}$ and the Fermi $k_{F}$ wavenumbers. Here, $k_{\ell}$ is defined by $\tilde{\xi} (k_{\ell})=0$, where $\tilde{\xi}(\mathbf{k})$ is defined in Eq.\ \eqref{eq:xiergdef}; and $k_{F}$ is defined by $f_{e}^{(0)} (k_{F}) = \frac{1}{2}$ and $p_{eh}^{(0)} (k_{F}) = 0$.
		Note that $k_{\ell} \neq k_{BCS}$, where the BCS gap wavenumber $k_{BCS}$ is defined as $2\tilde{E}(k_{BCS}) = \tilde{E}_{gap}^{pair}$ (see Eqs.\ \eqref{equ:E-single-particle-open} and \eqref{eq:def-epairgap}).
		(b) An evenly-spaced sampling of UU continuum modes, over a portion of the energy range at which the $\tilde{E}(k)$ band is doubly degenerate with respect to $k$.
		Close to $k_{BCS}$, the eigenfunctions have two peaks. These peaks occur at pairs of $k$ values, $k_{1}$ and $k_{2}$, for which $\tilde{E} (k_1) = \tilde{E} (k_2)$. For eigenenergies greater than the band where  $\tilde{E}(k)$ is $k$-degenerate, the eigenfunction magnitudes are simple $k$-peaks.
		The LL modes are identical, except for the interchanges $\delta P \leftrightarrow \delta P^{\ast}$ and $\delta E \leftrightarrow \delta E^{\ast}$, where $(\delta P, \delta E)$ and $(\delta P^{\ast}, \delta E^{\ast})$ correspond to $\vec{X}$ and $\vec{Y}$ in Eq.\ \eqref{eq:eigveccomp}, respectively. Their $\delta f$ components ($\vec{Z}$ in Eq.\ \eqref{eq:eigveccomp}) are identical. $|\delta P|$ and $|\delta E|$ are greater than $|\delta P^{\ast}|$ and $|\delta E^{\ast}|$ in the UU continuum, while per the interchange symmetry, this is reversed for LL. The THz response matrix $M$ eigenfunctions are very similar, except they lack the $\delta E$ and $\delta E^{\ast}$ components.}
	\label{fig:eigenvec}
\end{figure*}

Figure \ref{fig:eigenvec} shows the eigenvectors for selected modes, with Fig. \ref{fig:eigenvec}(a) showing collective modes and Fig. \ref{fig:eigenvec}(b) continuum states.

\begin{figure}
	\centering
		\includegraphics{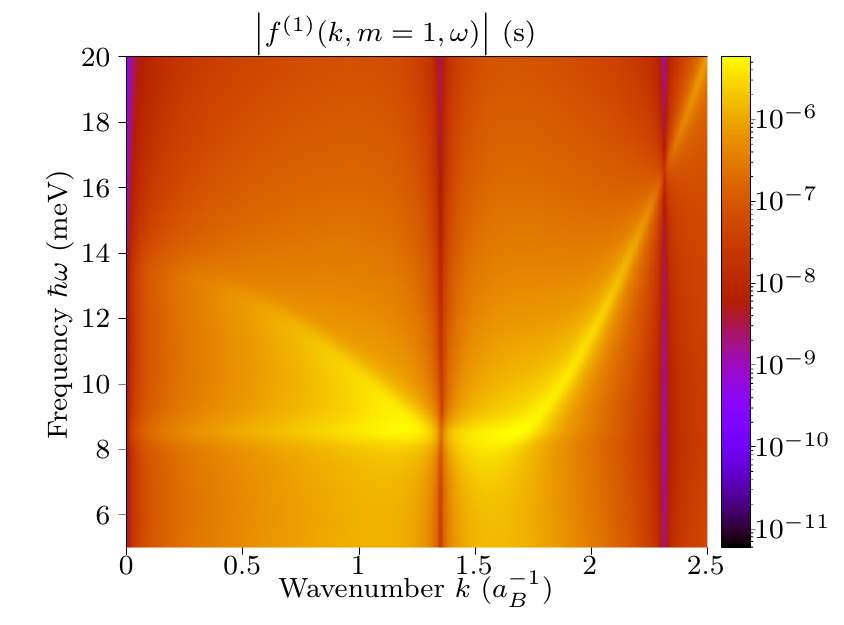}
	\caption{
		(Color online.)
		First-order density (carrier distribution function) response $|f_{e}^{(1)} (k,m=1,\omega)|$.
		The curved (yellow) peak location line corresponds to vertical transitions within, say, the conduction bands shown in Fig. \protect\ref{fig:bandstruct},
		and the horizontal peak line corresponds to the T modes.
		%
		%The branches correspond to $2\tilde{E}_{k}$.
		%
		%There is also a branch from $k=0$ to the second intersection on $2\tilde{E}_{k}$ at $\approx 2\tilde{E}_{k_{\ell}}$,
		% corresponding to the T modes.
		%%	For all $k$ there is a peak at $\omega = 0$, as is expected from the $1/\omega$
		%%   factor that comes from the THz source vector.
		The $f^{(1)}=0$ line at 1.35 (2.4) $a_B^{-1}$ is
		at $k_{\ell}$ ($k_{F}$), defined in the caption to Fig.\ \ref{fig:eigenvec}.
		%and the Fermi wave vector.
		%$k_{F}$.
	}
	\label{fig:sm1233_f1_abs_clrmap_no0}
\end{figure}

\begin{figure}
	\centering
	\includegraphics{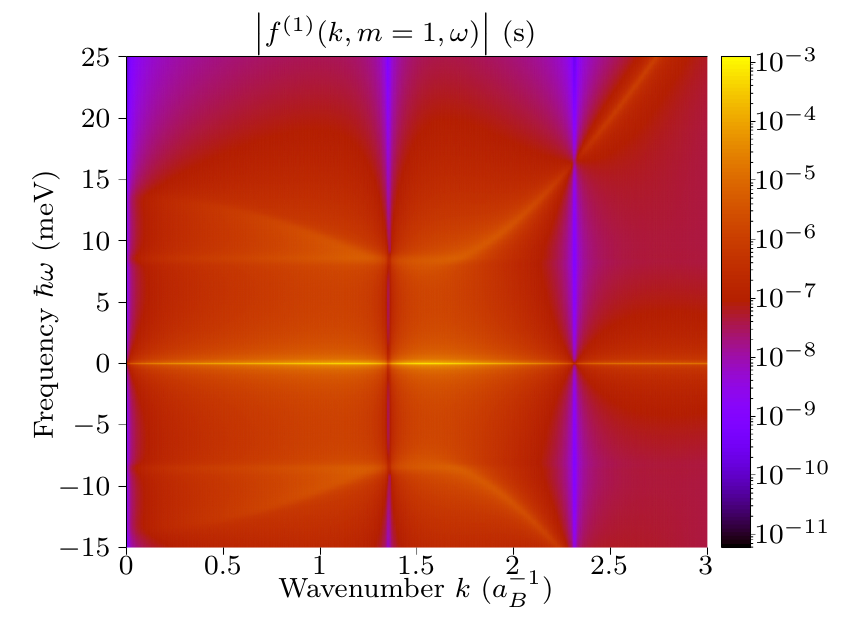}
	\caption{Color map of $|f_{e}^{(1)} (k,m=1,\omega)|$.
		Maxima occur at $\pm 2\tilde{E}_{k}$, corresponding to the structure in the eigenvalues $\Re \varepsilon_{k}$; at $\omega = 0$ for all $k$, as is expected from the $1/\omega$ factor present in the THz source vector $\vec{s}$; and at $\hbar\omega \approx 2\tilde{E}_{k_{\ell}}$, from $k=0$ to the second intersection with $2\tilde{E}_{k}$, corresponding to the T$_0$ and T$_1$ resonance.
		There are two minima, at $k = k_{\ell}$ and at $k = k_{F}$ for all $\omega$, which correspond to the minima in $|p_{eh}^{(0)} (k)|$.}
	\label{fig:f1_abs_clrmap}
\end{figure}

\begin{figure}
	\centering
		\includegraphics{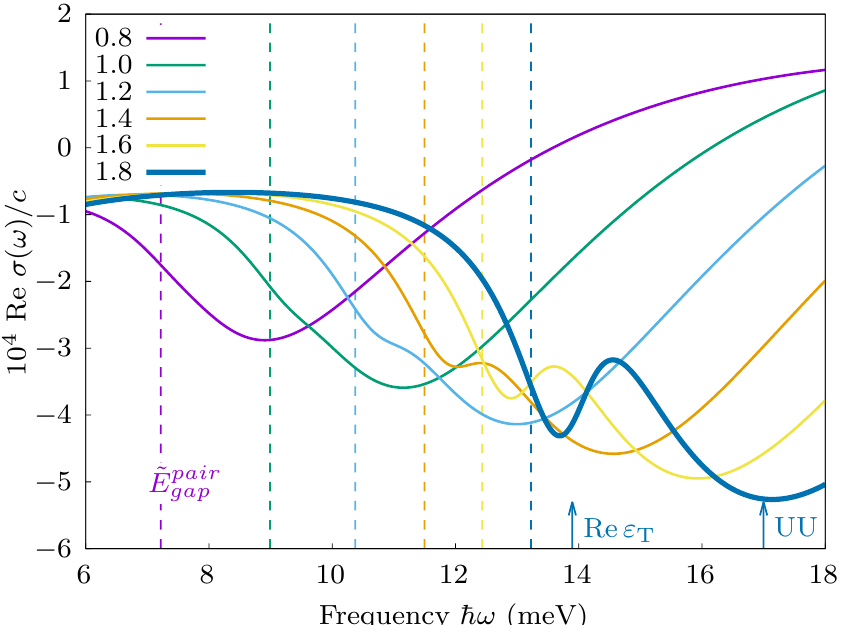}
	\caption{
		The real part of the total conductivity, which is proportional to the absorptivity $A$, as a function of frequency. The key is pump density in units of inverse effective Bohr radius squared $a_{B}^{-2}$. All pump densities are above threshold, which is at $0.41/a_B^2$.
		Here, $\gamma = 0.5$ meV,
		and  $\gamma_f = 1.4001$ meV.
		The dashed lines correspond to the frequency of the BCS-like gap, $\tilde{E}_{gap}^{pair}$.
		For the highest pump density, the frequencies of the T-modes and of the gain peak corresponding to the UU-continuum are indicated.
	}
	\label{fig:sm1261_cond_re_dens}
\end{figure}

   An important consequence of Fig. \ref{fig:lineigen} is the fact that both interband  and intraband spectra show a BCS-like gap. In both cases, the gap is not only that between the frequency of the order parameter and excited continuum states (at the hook-like feature), but also involves collective modes, M and T, in the vicinity of the continuum gap. These modes stem from the strong Coulomb interaction and are not present in a photon laser, where Coulomb interactions are negligible \cite{spotnitz-etal.2021}.
    In particular, Fig. \ref{fig:lineigen}a predicts the possibility to experimentally observe the BCS-like gap using THz radiation (more details below, cf. Fig. \ref{fig:sm1261_cond_re_dens}).

   To further analyze the occurrence of collective modes due to the many-particle Coulomb interaction, it is helpful to look at the first-order (in the probe) modification of the carrier distribution, $f^{(1)}$,
as a function of wave vector and frequency, Fig.\ \ref{fig:sm1233_f1_abs_clrmap_no0}. In addition to the light-induced band from the single-particle spectrum, we see a horizontal line at about 8.5 meV that is due to the excitation of the collective T modes as shown
   in Fig.\ \ref{fig:lineigen}.
   Fig.\ \ref{fig:f1_abs_clrmap} provides a larger-scale view of  \ref{fig:sm1233_f1_abs_clrmap_no0}.

\begin{figure}
	\centering
		\includegraphics[width=0.45 \textwidth]{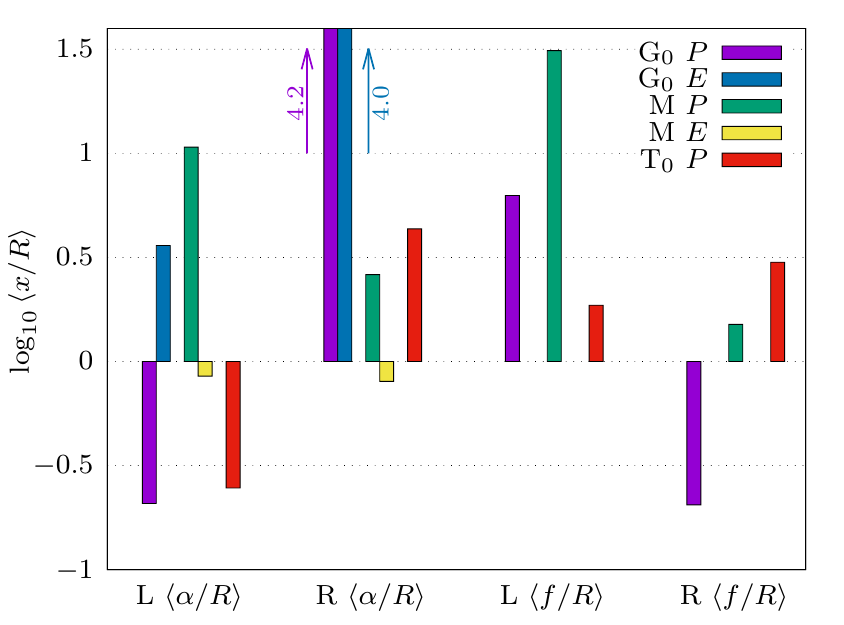}
	\caption{
		(Color online.)
		Ratios of the arc length $\alpha$ and density $f$ to the amplitude components $R$ of selected collective modes (Goldstone, M and T according to
		Fig. \protect\ref{fig:lineigen}) for right (R) and left (L) eigenvectors with averaging over $k$ as
		described in text.
	}
	\label{fig:ampl_phas_box_log}
\end{figure}

\begin{figure*}
	\centering
	\includegraphics{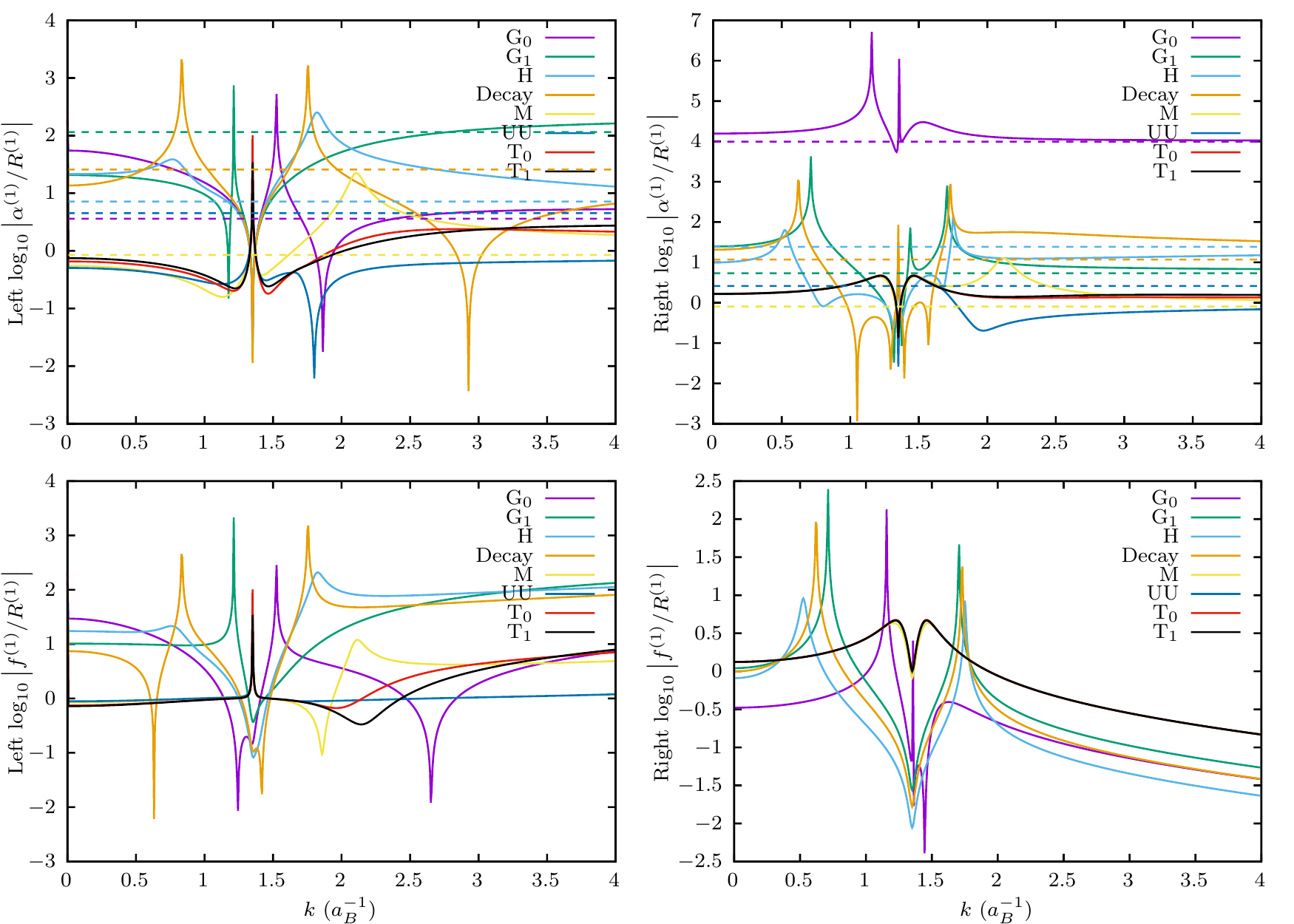}
	\caption{The base-ten logarithm of the arc length-to-amplitude $\left|\alpha^{(1)} / R^{(1)} \right|$ and density-to-amplitude $\left|f^{(1)} / R^{(1)} \right|$ ratios as a function of wavenumber $k$ for the left and right eigenvectors of selected modes. In the plots of $\left|\alpha^{(1)} / R^{(1)} \right|$ for the optical-probe eigenvectors, dashed lines denote the $\delta E$ component, and solid lines denote the $\delta P$ component. The THz-probe modes have no associated $\delta E$ component, and so only the $\delta P$ component of the $\left|\alpha^{(1)} / R^{(1)} \right|$ ratio is plotted for T$_0$ and T$_1$.
		The $\delta E$ eigenvector components have no defined $k$-dependence, and so their $\left|\alpha^{(1)} / R^{(1)} \right|$ ratios are plotted as constant in $k$. The figure shows the $\delta E$ ratios are the asymptotic limit for high $k$ of the $\delta P$ ratios.
		The UU and Decay modes are merely example modes taken from their respective continua.
		All modes except T$_0$ and T$_1$ are taken from the optical probe response matrix.
		Many ratios exhibit extrema near $k_{\ell} = 1.35 a_{B}^{-1}$.
		The main outlier in the set is that the $\left|\alpha^{(1)} / R^{(1)} \right|$ ratios of the right eigenvectors show the undamped Goldstone mode G$_0$ to clearly be a phase mode.}
	\label{fig:ampl_phas_lograt}
\end{figure*}

\subsection{Mode characterization}
\label{subsec:mode-characterization}

   Conventional discussions of fluctuation modes focus on whether the modes are phase or amplitude modes \cite{varma.2002}.
   However, in our general theory  the fluctuations also involve density fluctuation.
     To analyse the fluctuation modes in terms of phase, amplitude, and density fluctuations, we write each interband variable (i.e., each $p(k,m)$ and $E$) as a complex number
     $z=Re^{i \phi}$, with $R^{(0)}e^{i \phi^{(0)}}$ the zeroth-order steady state solution.
     The phase $\phi$ is varied at first order in the perturbation, leading to a first-order arc length variation  $\alpha^{(1)} = R^{(0)}  \phi^{(1)}$.
     The first-order (in the external probe) amplitude modulation is denoted by $R^{(1)}$.
 %
 %    The first-order modulation of the distribution function is called $f^{(1)}$.
 %
     In order to avoid ambiguities stemming from the separate normalization of each left (L) and right (R) eigenvector, we show the ratios
     $\alpha^{(1)}/R^{(1)}$ and $f^{(1)}/R^{(1)}$ for both L and R.

 %    We note that in simple models without density fluctuations and one-component order parameters,
 %     a pure phase mode (Goldstone) has no amplitude fluctuation in the right eigenvector, $R^{(1)}=0$, hence  $\alpha^{(1)}/R^{(1)} = \infty$,
 %     and in models with at least two dynamic variable where a pure
 %     amplitude mode (Higgs) is possible, that mode has no phase fluctuation, $\alpha^{(1)}/R^{(1)} = 0$.

  %    In our many-particle approach, the label Goldstone (Higgs) mode requires all components of $P(k,m)$ and $E$ to be Goldstone (Higgs) modes.
  %    The presence of density fluctuation does not affect this labeling. Furthermore, the labelling is based on the right eigenvectors. The left eigenvectors %are needed to construct the response function (intraband current) and are in general different from the right eigenvectors. Broadly speaking,
  %    if the right eigenvector contains a $u$-component (where $u$ stands for phase, amplitude, density) and the left eigenvector contains a $v$-component,
  %    then a external source of $v$-character can induce a fluctuation of $u$-character.
  %
      In our many-particle approach, the right eigenvector of a Goldstone mode has non-zero phase fluctuations (in $p(k)$ and $E$) but zero amplitude and density fluctuations for each $k$. The right eigenvector of a Higgs mode has zero phase fluctuations for each $k$. Both left and right eigenvectors are needed to construct the response function. The fluctuation matrix being non-Hermitian, the left and right eigenvectors for the same eigenvalue can be quite different. It is hence possible that the response function may mix the three subsets of amplitude, phase and density.
      For example, an external amplitude fluctuation can create a phase mode if the left eigenvector has an amplitude component and the right eigenvector a phase component.
     See Sec.\ \ref{sec:phasampresp} for more details.

      Figure \ref{fig:ampl_phas_box_log} summarizes our results as means. Rather than showing the $k$-resolved results, we show averages over $k$.
      In the horizontal axis, L or R denotes the ratio for the left or right eigenvectors, respectively. $R$, $\alpha$, and $f$ are the amplitude, arc-length, and density components of the eigenvectors, and  $\left\langle  \alpha/R \right\rangle$ and $\left\langle  f/R \right\rangle$ denote whether the ratio to $R^{(1)}$ is of the arc length or the density, respectively. In the vertical axis label, the expression $\left\langle  x/R \right\rangle$ denotes
		\begin{equation}
\left\langle  x/R \right\rangle \equiv
			\begin{cases}
				\frac{1}{N_{k}} \sum_{i=1}^{N_{k}} \left\vert x^{(1)}(k_{i})/R^{(1)}(k_{i})  \right\vert , & \text{for} \ P^{(1)} \\
				\left\vert x^{(1)}/R^{(1)} \right\vert, & \text{for} \ E^{(1)} .
			\end{cases}
\end{equation}
%
%
%      %
%	 \begin{equation}
%\left\langle  x/R \right\rangle \equiv
%			\begin{cases}
%				\frac{1}{N_{k}} \sum_{i=1}^{N_{k}} \left\vert x^{(1)}/R^{(1)}  \right\vert (k_{i}), & \text{for} \ P^{(1)} \\
%				\left\vert x^{(1)}/R^{(1)} \right\vert, & \text{for} \ E^{(1)} .
%			\end{cases}
%\end{equation}
%%
%
   Figure \ref{fig:ampl_phas_lograt} provides the underlying $k$-resolved data.

     Figure \ref{fig:ampl_phas_box_log} gives the justification to call $G_0$ a Goldstone (phase) mode, as the numerical result yields the phase component (presented as arc length) of both the interband polarization and the cavity field four orders of magnitude larger than the amplitude and density fluctuations. On the other hand, the collective modes M in the interband response and T in the THz response are mixed modes since all ratios (arc length/amplitude, density/amplitude) are approximately of order 1;  they are neither Higgs nor Goldstone modes.
     We do not observe pure Higgs modes.

\subsection{Spectroscopic observables}
\label{sec:spectroscopic-observables}

\begin{figure}
	\centering
	\includegraphics{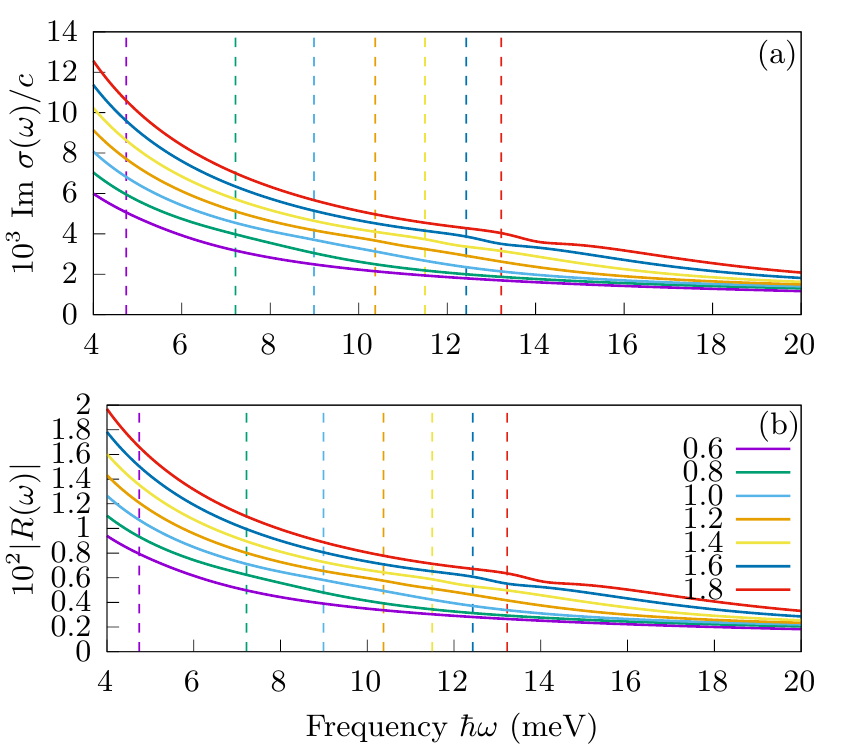}
	\caption{(a) The imaginary part of the THz-domain conductivity, $10^{3} \Im \sigma_{T}(\omega)/c$. (b) The magnitude of the reflection coefficient amplitude $|R(\omega)|$, as a percentage. Note that $R$ is the ratio of reflected to incident amplitudes, while $|R|^2$ is the ratio of intensities.
		Both (a) and (b) are plotted against the frequency $\hbar\omega$ (meV).
		The key is the pump density $n_{p}$ in units of inverse effective Bohr radius squared $a_{B}^{-2}$.
		The dashed lines denote the BCS gaps $\tilde{E}_{gap}^{pair}$ for the corresponding pump densities.
		Both $\Im \sigma(\omega)$ and $|R(\omega)|$ are dominated by their $\omega^{-1}$ dependence, coming from the diamagnetic part of the conductivity, but some features can be seen in the vicinity of $\tilde{E}_{gap}^{pair}$, particularly at the highest $n_{p}$. Any features around the BCS gap result from the paramagnetic part of the conductivity.}
	\label{fig:cond_im_r_np}
\end{figure}

\begin{figure}
	\centering
	\includegraphics{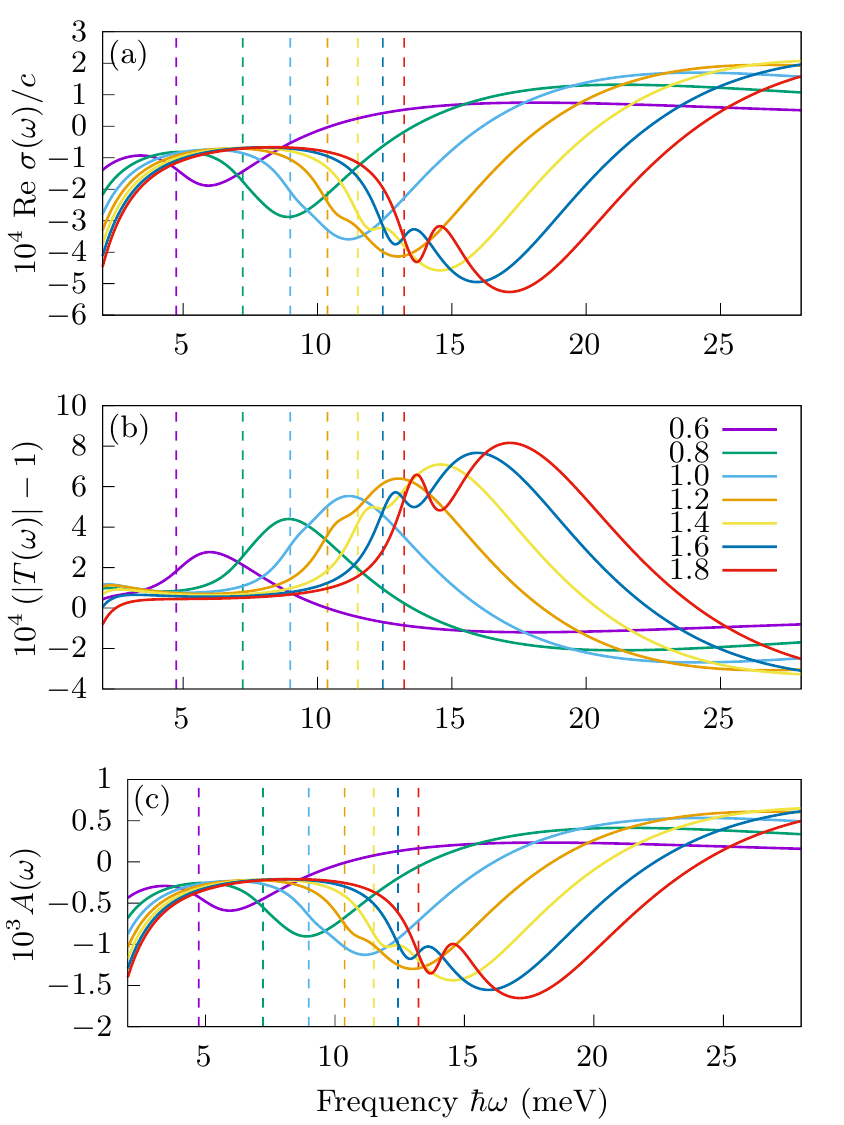}
	\caption{Plotted versus THz probe frequency $\hbar\omega$ (meV): (a) Real THz-frequency conductivity $10^{4} \Re \sigma_{T}(\omega)/c$. (b) Transmission coefficient $10^{4}\left(|T(\omega)|-1\right)$. (c) Absorptivity $10^{3} A(\omega)$.
		The key is the pump density $n_{p}$ in units of $a_{B}^{-2}$.
		The dashed lines denote the BCS gaps $\tilde{E}_{gap}^{pair}$ for the respective pump densities.
		Note that $T$ is the ratio of transmitted to incident amplitudes, while $|T|^2$ is the ratio of intensities. $A$ is the fraction of incident intensity which is absorbed. $A <0$ denotes gain.
		At higher pump densities, two minima may be seen in $\Re \sigma$ and $A$, in the vicinity of the BCS gap. The first valley corresponds to the frequency of the T$_0$ and T$_1$ modes.
		As the pump density drops below that required for the emergence of the T modes, the second minimum spreads out to become a single broad minimum.
	}
	\label{fig:cond_re_at_np}
\end{figure}

\begin{figure}
	\centering
	\includegraphics{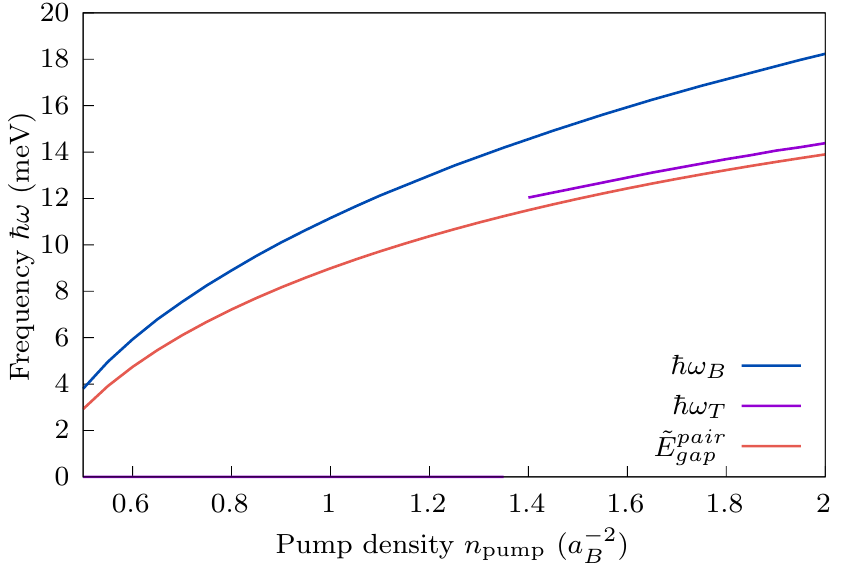}
	\caption{Comparison of the frequencies $\hbar\omega$ (meV) of the BCS-like gap $\tilde{E}^{pair}_{gap}$ and of the minima $\omega_{B}$ and $\omega_{T}$ of the THz-domain conductivity $\Re \sigma_{T}(\omega)$, for varying pump density $n_{p}$.
		The light red line denotes the magnitude of the BCS-like gap, $\tilde{E}^{pair}_{gap}$.
		The dark blue line shows the frequency $\hbar\omega_{B}$ for which the real part of the conductivity $\Re\sigma(\omega)$ is a minimum; that is, $\hbar\omega_{B}$ is defined such that $\Re\sigma(\omega_B)=\min \Re\sigma(\omega)$. For pump densities at which the T modes can be discerned, the $\hbar\omega_{B}$ minimum is the broader, deeper, higher-frequency conductivity minimum.
		The purple line tracks the frequency $\omega_{T}$ of the first, smaller minimum in $\Re \sigma$, corresponding to the T modes. This is assigned $\omega_{T}=0$ when the peak-finding algorithm cannot find this minimum, i.e., for pump densities below the emergence of the T modes.
		The plot shows that the BCS-like gap $\tilde{E}^{pair}_{gap}$ is closely tracked by the gain maximum, over a wide range of pump densities, and in particular by the frequency $\hbar\omega_{T}$ of the T mode $\Re \sigma$ minimum.
	}
	\label{fig:bcs_gap_dens}
\end{figure}

Finally, we use the eigenvectors and eigenvalues to obtain the intraband conductivity, the real part of which is proportional to the THz absorption. Figure
\ref{fig:sm1261_cond_re_dens} shows extrema in the conductivity that approximately scale with the  BCS-like gap, hence predicting a THz-based observation of the gap. At low pump densities, we have one broad extremum stemming mostly from the continuum states, which, in combination, lead to a broad THz gain band. At high pump densities (here $1.4 a_{B}^{-2}$), a narrow line becomes visible. This stems from the two T-modes in Fig.\ \ref{fig:lineigen}.
They too yield a region of THz gain, but this gain peak  becomes a narrow resonance since
the imaginary part of their eigenvalues $\frac{1}{2} (\Im \varepsilon_{T_0} + \Im \varepsilon_{T_1})$
approaches the dephasing (specifically, $-\frac{1}{2} (\gamma + \gamma_f)$; see  Fig.\ \ref{fig:thz_eigval_tmode_track}),
which can in principle be made very small.
As the pump density is increased or the interband dephasing rate is decreased (beyond what is plotted here), the T-modes yield a dispersive-like feature in $\Re \sigma_{T}$ and a Lorentzian feature in $\Im \sigma_{T}$, with increasing height-to-width ratios.

Figure \ref{fig:cond_im_r_np} provides corresponding data of the imaginary part.
Figure \ref{fig:cond_re_at_np} provides data analogous to Fig.\ \ref{fig:sm1261_cond_re_dens}, but for a larger selection of pump densities and also showing, in addition to the real part of the THz conductivity, plots of the transmission and absorption.
 Fig.\ \ref{fig:sm1261_cond_re_dens} shows that the extrema in the THz conductivity track the polaritonic BCS gap (shown as vertical dashed lines in that figure).
This tracking behavior is made more apparent and summarized in
 Fig.\ \ref{fig:bcs_gap_dens}.

\section{Conclusion}

In summary, we have shown that, in a polariton laser operating in the polaritonic BCS regime,
fluctuations or external probes far detuned from the laser frequency $\omega_\ell$, i.e., the frequency of the order parameter, can induce fluctuation modes that are different from those induced by fluctuations or probes close to $\omega_\ell$.
In addition to the importance of their existence, finding the characteristic properties of these modes can help with future experimental identification.
The fluctuation modes in this new set have an orbital angular momentum different from that of the order parameter; they contain spectral continua as well as collective modes, but do not contain the Goldstone modes.
The collective modes are shown not to be pure amplitude (Higgs) modes.
All modes can contribute to THz gain, including collective modes T$_i$ with frequencies close to the BCS-like gap.
The THz gain resonance from these T modes can, at least in principle, be made arbitrarily narrow, as their
width-to-height ratio decreases with the interband dephasing rate and with increasing pump density.
In this limit, the T$_0$ modes may also become unstable; this warrants further investigation.
The THz gain mechanism presented here is novel, but experimental verification is needed and its efficiency should be compared with that of existing THz emitters.
Future comparison of the modes found here with Bardasis-Schrieffer polaritons defined in thermal equilibrium systems \cite{bardasis-schrieffer.1961,sun-millis.2020} is desirable, as it would enhance the understanding of fluctuations in lasers, being examples of condensed open-pumped-dissipative systems.
Furthermore, as discussed in Ref.\ \cite{hu-etal.2021}, the experimental observation of the new modes can help solidify evidence for the polaritonic BCS states.

\begin{acknowledgments}

We gratefully acknowledge useful discussions with Hui Deng, University of Michigan;
financial support from the NSF under grant number DMR 1839570;
and the use of High Performance Computing (HPC) resources supported by the University of Arizona.

\end{acknowledgments}

\appendix

\section{Symmetry properties of eigenvalues, eigenvectors, and source vectors}
\label{appx:vector-symmetries}

%The laser linear response equation has the general form of . The matrix
The matrix $M$ in the linear response Eq.\ \eqref{eq:delta-x-dot} has the following structure
%We arrange the response field variables in each angular momentum channel as a vector in the following way:
%\begin{equation}	%\vec{x}(t)=\left(\begin{matrix}{\vec{\xi}}^{(1)}(t)\\{\vec{\xi}}^{(1)\ast}(t)\\{\vec{f}}^{(1)}(t)\\\end{matrix}\right)\quad,\quad\vec{\xi}(t)=\left(\begin{matrix}{\vec{p}}_{eh}^{(1)}(t)\\E^{(1)}(t)\\\end{matrix}\right)
%	\label{eq:xtblockform}
%\end{equation}
%There are $N_k$ radial $k$ points, and the vectors $\vec{p}_{eh}^{(1)}(t), \vec{p}_{eh}^{(1)\ast}(t), \vec{f}^{(1)}(t)$ are the values of the variables at these $k$ points. We have grouped $\vec{p}_{eh}^{(1)}(t)$ and $E^{(1)}(t)$ into the vector $\vec{\xi}^{(1)}(t)$ for convenience. The matrix $M$ is $N\times\ N$ with $N=3N_k+2$. $M$ and the source $\vec{s}(t)$ the following structure
\begin{equation}
	M=\left(\begin{matrix}A&B&\widetilde{C}\\-B^{\ast}&-A^{\ast}&-{\widetilde{C}}^{\ast}\\\widetilde{D}&-{\widetilde{D}}^{\ast}&G\\\end{matrix}\right)
	%\quad,\quad\vec{s}(t)=\left(\begin{matrix}{\vec{s}}_\xi(t)\\-{\vec{s}}_\xi^{\ast}(t)\\{\vec{s}}_f(t)\\\end{matrix}\right)\label{eq:Mblockform}
\end{equation}
where $A$, $B$, $\widetilde{C}$, $\widetilde{D}$, and $G$ are $N_k \times N_k$ sub-matrices. $A$, $-A^{\ast}$, and $G$ are sub-matrices in the ${\vec{p}}^{\, (1)}_{e h}$, ${\vec{p}}^{\, (1) \ast}_{e h}$, and ${\vec{f}}^{\, (1)}_e$ blocks respectively. $B$ couples ${\vec{p}}^{\, (1)\ast}_{e h}(t)$ to ${\vec{p}}^{\, (1)}_{e h}(t)$, $\widetilde{C}$ couples ${\vec{f}}^{\, (1)}_e(t)$ to ${\vec{p}}^{\, (1)}_{e h}(t)$, and $\tilde{D}$ couples ${\vec{p}}^{\, (1)}_{e h}(t)$ to ${\vec{f}}^{\, (1)}_e(t)$, etc.
For the case where $E_{\ell\lambda}^{(1)} \neq 0$, such as with an optical (interband) probe, $E_{\ell\lambda}^{(1)}$ can be treated as the $(N_{k}+1)^{\mathrm{th}}$ component of the $\vec{p}_{eh}^{\, (1)}$ vector, and these theorems still apply, with certain dimensions changed from $N_{k}$ to $N_{k} +1$.
The block matrix structure leads to some symmetries among the eigenvalues and eigenvectors, including the following.

%\begin{thm}
\noindent
\textit{Claim 1}:
Suppose $\lambda$ is an eigenvalue and $\vec{\zeta} \equiv \begin{pmatrix}\vec{X}\\\vec{Y}\\\vec{Z} \end{pmatrix}$
is the corresponding right (left) eigenvector. If the sub-matrix $G$ is purely imaginary (which, in this paper, it is), then $-\lambda^{\ast}$ is also an eigenvalue with right (left) eigenvector $\vec{\zeta}^{\, \prime} \equiv \begin{pmatrix}{\vec{Y}}^{\ast}
	\\ {\vec{X}}^{\ast} \\ {\vec{Z}}^{\ast} \end{pmatrix}.$
\label{thm:eigvalnegconjgsym}
%\end{thm}

\begin{proof}
	The equations for eigenvalue $\lambda$ and its right eigenvector are
	\begin{IEEEeqnarray}{rCl}
		A\vec{X}+B\vec{Y}+\widetilde{C}\vec{Z} &=& \lambda\vec{X} \label{eq:blockmatrow1} \\
		-B^{\ast}\vec{X}-A^{\ast}\vec{Y}-{\widetilde{C}}^{\ast}\vec{Z} &=& \lambda\vec{Y} \label{eq:blockmatrow2} \\
		\widetilde{D}\vec{X}-{\widetilde{D}}^{\ast}\vec{Y}+G\vec{Z} &=& \lambda\vec{Z} \label{eq:blockmatrow3}
	\end{IEEEeqnarray}
	Take the complex conjugate of each equation and multiply through by $-1$. Then switch the order of the first two equations to get
	\begin{IEEEeqnarray}{rCl}
		A{\vec{Y}}^{\ast}+B{\vec{X}}^{\ast}+\widetilde{C}{\vec{Z}}^{\ast} &=& -\lambda^{\ast}{\vec{Y}}^{\ast} \label{eq:blockmatprimerow1} \\
		-B^{\ast}{\vec{Y}}^{\ast}-A^{\ast}{\vec{X}}^{\ast}-{\widetilde{C}}^{\ast}{\vec{Z}}^{\ast} &=& -\lambda^{\ast}{\vec{X}}^{\ast} \label{eq:blockmatprimerow2} \\
		\widetilde{D}{\vec{Y}}^{\ast}-{\widetilde{D}}^{\ast}{\vec{X}}^{\ast}-G^{\ast}{\vec{Z}}^{\ast} &=& -\lambda^{\ast}{\vec{Z}}^{\ast} \label{eq:blockmatprimerow3}
	\end{IEEEeqnarray}
	If $G$ is purely imaginary, $-G^{\ast}=G$, and comparing Eqs.\ \eqref{eq:blockmatprimerow1}--\eqref{eq:blockmatprimerow3} with Eqs.\ \eqref{eq:blockmatrow1}--\eqref{eq:blockmatrow3} shows $-\lambda^{\ast}$ is an eigenvalue with right eigenvector ${\vec{\zeta}}^{\, \prime}$.  If $\mathrm{Re}\lambda=0$, then $-\lambda^{\ast}=\lambda$, and we have $\vec{X}={\vec{Y}}^{\ast}$ and ${\vec{Z}}^{\ast}=\vec{Z}$.

	The transpose $M^T$ has the same block structure symmetry relevant to the proof as $M$. So the claim is also valid for the left eigenvector. Explicitly, if $\vec{\zeta}$ is the left eigenvector, the transpose eigenvalue equation is
	\begin{IEEEeqnarray}{CCrCl}
		%{C+c+rCl}
		\vec{\zeta}^{\, \dagger} M=\lambda \vec{\zeta}^{\, \dagger} & \ \Leftrightarrow \  & \vec{X}^{\dagger} A-\vec{Y}^{\dagger} B^{*}+\vec{Z}^{\dagger} \tilde{D}&=& \lambda \vec{X}^{\dagger} \\
		& &\vec{X}^{\dagger} B-\vec{Y}^{\dagger} A^{*}-\vec{Z}^{\dagger} \tilde{D}^{*}&=&\lambda \vec{Y}^{\dagger} \\
		& & \vec{X}^{\dagger} \tilde{C}-\vec{Y}^{\dagger} \tilde{C}^{*}+\vec{Z}^{\dagger} G &=& \lambda \vec{Z}^{\dagger}
	\end{IEEEeqnarray}
	Repeating the arguments above for the right eigenvector shows that $\vec{\zeta}^{\, \prime}$ is the corresponding left eigenvector for the eigenvalue $- \lambda^{\ast}$.
\end{proof}

Of course, an eigenvector is defined only up to an overall constant. So the eigenvectors of $\lambda$ and $-\lambda^{\ast}$ obtained, e.g., from computation, may not automatically satisfy the above relation between $\vec{\zeta}$ and ${\vec{\zeta}}^{\, \prime}$. Multiplication by an appropriate overall constant would make the relation valid. The same caveat applies to the relation between the eigenvector elements when $\mathrm{Re}\lambda=0$.

\noindent
%\begin{thm}
\textit{Claim 2}. Consider the block representation Eq.\ \eqref{eq:somblockform} of the frequency-domain source vector $\vec{s} (m , \omega)$, and let $\vec{z}_j (|m|)$ and $\vec{z}^{\, \prime}_j (|m|)$ be the (left) eigenvectors corresponding to a certain pair of eigenvalues $(\lambda_j, -\lambda^{\ast}_j)$ and have the block forms as laid out above in Claim 1.  If $\vec{s}_{ f}(m, \omega)=-\vec{s}_{f}^{\, *}\left(-m, -\omega\right)$, (which, in this paper, follows trivially from $\vec{s}_{f}=0$,) then
\begin{equation}
	-\left(\vec{z}^{T}_{j}(|m|) \vec{s}\left(-m, -\omega \right)\right)^{*}=\vec{z}_{j}^{\, \prime T}(|m|) \vec{s}(m, w)
\end{equation}
%\end{thm}
\begin{proof}
	Writing $\vec{z}_{j}(|m|)=\left(\begin{array}{c}\vec{X} \\ \vec{Y} \\ \vec{Z}\end{array}\right), \vec{z}_{j}^{\, \prime}(|m|)=\left(\begin{array}{c}\vec{Y}^{\ast} \\ \vec{X}^{*} \\ \vec{Z}^{*}\end{array}\right)$, and $\vec{s}(m, \omega)$ as above, we have
	\begin{widetext}
		\begin{IEEEeqnarray}{rCl}
			\vec{z}^{\, T}_{j}(|m|) \vec{s}(m, w) &=& \vec{X}^{T} \vec{s}_{p}(m, w) -\vec{Y}^{T} \vec{s}_{p}^{\, \ast}(-m,-w) +\vec{Z}^{T} \vec{s}_{f}(m, w) \label{eq:thm3proofeq1} \\
			-\left(\vec{z}^{\, T}_{j}(|m|) \vec{s}\left(-m, - \omega\right)\right)^{\ast} &=& -\vec{X}^{\dag} \vec{s}_{p}^{\, *}\left(-m, -\omega\right)+\vec{Y}^{\dag} \vec{s}_{p}(m, \omega) -\vec{Z}^{\dag} \vec{s}_{f}^{\, \ast}\left(-m, -\omega\right) \label{eq:thm3proofeq2} \\
			\vec{z}_{j}^{\, \prime T}(|m|) \vec{s}(m, \omega) &=& \vec{Y}^{\dag} \vec{s}_{p}(m, \omega)-\vec{X}^{\dag} \vec{s}_{p}^{\, \ast}\left(-m, -\omega\right)+\vec{Z}^{\dag} \vec{s}_{f}\left(m, \omega\right) \label{eq:thm3proofeq3}
		\end{IEEEeqnarray}
	\end{widetext}
	If $\vec{s}_{f}\left(m, \omega\right)=-\vec{s}_{f}^{\, \ast}\left(-m,-\omega\right)$, then Eqs.\ \eqref{eq:thm3proofeq2} \& \eqref{eq:thm3proofeq3} are equal.
\end{proof}
Claim 2 enables simplification of the coefficients $c_n(\omega)$ in Eq.\ \eqref{eq:linrespvecomeigvecexp}.

%\section{Simplification of the Coulomb Matrix Elements $V_{k,k^{\prime}}^{m,m^{\prime}}$}
% Properties of the
\section{The Angular Momentum Coulomb Matrix Elements}
% This expression is not allwoed in a section title hyperref:
 %$V_{k,k^{\prime}}^{m}$}
\label{appx:coulmel}

There are multiple ways to derive and write Eq.\ \eqref{eq:vkkpmfin} for the Coulomb matrix elements in the angular momentum basis, which is used in Eqs.\ \eqref{eq:delta-P-dot-ang}--\eqref{eq:delta-E-star-dot-ang}.
The method used in Eq.\ \eqref{eq:vkkpmfin} is to expand $V_{|\mathbf{k}-\mathbf{k}'|}^{c}$ as a Fourier series in the relative reciprocal space angle $\theta_{k}-\theta_{k}'$.
As stated below Eq.\ \eqref{eq:vkkpmfin}, the Coulomb matrix elements notably do not couple different angular momentum components.
The Coulomb interaction is independent of the absolute angle in real or reciprocal space, so by Noether's theorem, it conserves angular momentum.
%$m$ and m' must be equal on the two sides of the equation.
%Since $V_{|\mathbf{k}-\mathbf{k}'|}^{c}$ is a periodic function of $\theta-\theta'$, it can be expanded as a Fourier series in $\theta - \theta'$.
Alternatively, that the Coulomb interaction between different angular harmonics is zero can be shown by explicit evaluation of the integral.

In substituting Eq.\ \eqref{eq:modeexpans} into Eqs.\ \eqref{eq:delta-P-dot}--\eqref{eq:delta-f-dot}, and then simplifying, one obtains the multi-variable integral for the Coulomb matrix elements
\begin{multline}
	V_{k,k^{\prime}}^{m,m^{\prime}} = \frac{2\pi e^{2}}{\epsilon_{b}} \int_{0}^{2\pi} \int_{0}^{2\pi} \frac{\mathrm{d} \phi \, \mathrm{d}\phi^{\prime}}{(2\pi)^{2}} \\
	\times \frac{e^{i(m\phi + m^{\prime} \phi^{\prime})}}{\sqrt{k^{2} + k^{\prime^{2}} - 2 k k^{\prime} \cos (\phi-\phi^{\prime})} + \kappa_{0}} \, .
\end{multline}
This can be simplified.
Using the change of variables $(\theta, \psi) = (\tfrac{1}{2}\phi -\tfrac{1}{2}\phi^{\prime},\tfrac{1}{2}\phi + \tfrac{1}{2}\phi^{\prime})$, this is
\begin{multline}
	V_{k,k^{\prime}}^{m,m^{\prime}} = \frac{e^{2}}{\pi \epsilon_{b}} \int_{-\pi}^{\pi} \mathrm{d} \theta \frac{e^{i(m-m^{\prime})\theta}}{\sqrt{k^{2} + k^{\prime^{2}} - 2 k k^{\prime} \cos 2\theta} + \kappa_{0}}  \\
	\times   \int_{|\theta|}^{2\pi - |\theta|} \mathrm{d}\psi \, e^{i(m+m^{\prime})\psi}
\end{multline}

For $m \neq -m^{\prime}$, the $\psi$ integral is evaluated to give
\begin{multline*}
	V_{k,k^{\prime}}^{m,m^{\prime}} = \frac{-2e^{2}}{\pi \epsilon_{b} (m+m^{\prime})} \\
	\times \int_{0}^{\pi} \mathrm{d} \theta \frac{\sin 2 m \theta + \sin 2 m^{\prime} \theta}{\sqrt{k^{2} + k^{\prime^{2}} - 2 k k^{\prime} \cos 2\theta} + \kappa_{0}}
\end{multline*}
The integrals over the two domains $[0,\frac{\pi}{2}]$ and $[\frac{\pi}{2},\pi]$ cancel each other, so that $V_{k,k^{\prime}}^{m,m^{\prime}} = 0$ for $m \neq -m^{\prime}$.

For $m = -m^{\prime}$, $V_{k,k^{\prime}}^{m,m^{\prime}}$ is nonzero, and so the matrix elements can be written as diagonal in $m$: $V_{k,k^{\prime}}^{m,m^{\prime}} = V_{k,k^{\prime}}^{m} \delta_{m,-m^{\prime}}$. Evaluating the $\psi$ integral gives the elements
\begin{equation}
	V_{k,k^{\prime}}^{m} = \frac{2e^{2}}{\epsilon_{b}} \int_{0}^{\pi} \mathrm{d} \theta \frac{\cos m \theta}{\sqrt{k^{2} + k^{\prime^{2}} - 2 k k^{\prime} \cos \theta} + \kappa_{0}} .
\end{equation}

For $\kappa_0 = 0$, $\mathbf{k} \neq \mathbf{k}'$, and $k(k') \neq 0$, Eq.\ \eqref{eq:vkkpmfin} may be represented in terms of elliptic integrals for the cases $m=0$\footnote{\cite{gradshteyn-ryzhik.2007} Eq.\ (2.571.5), p 179.} and $m=1$\footnote{\cite{gradshteyn-ryzhik.2007} Eq.\ (2.571.7), p 180.} as
%of the first kind, for the cases $m=0$ and $m=1$, which are the only cases which appear in the THz response, anyway. Elliptic integrals are not elementary functions, but standard algorithms exist for their evaluation, so this may be of some use. [See Numerical Recipes, Fortran 77, \S 6.11, p 254.]
%For $\kappa_0 = 0$, $\mathbf{k}\neq \mathbf{k}'$, $k \neq 0$, and $k' \neq 0$; defining $a \equiv k^2 + k'^2$, $b \equiv 2 k k'$, $c \equiv \frac{4e^{2}}{\epsilon_b}$, and $r \equiv \sqrt{\frac{2b}{a+b}}$,
\begin{IEEEeqnarray}{rCl}
	V_{k,k'}^{m=0} &=& \frac{c}{\sqrt{a+b}} K(r), \\
	%	\cite{[{}][{, Eq.\ 2.571.5, p 179.}]{gradshteyn-ryzhik.2007}}
%	\quad
	V_{k,k'}^{m=1} &=& \frac{c}{b\sqrt{a+b}} \left\{(b-a) \Pi\left(r^{2},r \right) + a K(r)\right\},
	%\cite{[{}][{, Eq.\ 2.571.7, p 180.}]{gradshteyn-ryzhik.2007}}
\end{IEEEeqnarray}
where $K(r)$
%= F(\frac{\pi}{2},r)$
is the complete elliptic integral of the first kind,\footnote{\cite{gradshteyn-ryzhik.2007} Eq.\ (8.112.1), p 860.}
%\cite{[{}][{, Eq.\ (8.112.1), p 860.}]{gradshteyn-ryzhik.2007}} % $F$ is the elliptic integral of the first kind, and
$\Pi$ is the complete elliptic integral of the third kind,\footnote{\cite{gradshteyn-ryzhik.2007} Eq.\ (8.111.4), p 860, with $\varphi = \frac{\pi}{2}$.}
%\cite{[{}][{, Eq.\ (8.111.4), p 860.}]{gradshteyn-ryzhik.2007}}
$a \equiv k^2 + k'^2$, $b \equiv 2 k k'$, $c \equiv \frac{4e^{2}}{\epsilon_b}$, and $r \equiv \sqrt{\frac{2b}{a+b}}$.
%
%For $k' = 0$, if $m \neq 0$, $V_{k,k'=0}^{m} = 0$; if $m=0$, $V_{k,k'=0}^{m} = \frac{2\pi e^2}{\epsilon_{b} (k + \kappa_0)}$.
% The integral
Eq.\ \eqref{eq:vkkpmfin} can be simply evaluated for $k'=0$, and if $\kappa_{0} \neq 0$, for $\mathbf{k}=\mathbf{k}'$:
\begin{IEEEeqnarray}{rCl}
	V_{k,k'=0}^{m} &=& \begin{cases}
		\frac{2\pi e^2}{\epsilon_{b} (k + \kappa_0)}, & \text{ if } m =0, \\
		0, & \text{ if } m \neq 0;
	\end{cases}
%	\qquad
	\\
	V_{\mathbf{k}=\mathbf{k}'}^{m} &=& \begin{cases}
		\frac{2\pi e^2}{\epsilon_{b}  \kappa_{0}}, & \text{ if } m =0, \\
		0, & \text{ if } m \neq 0.
	\end{cases}
\end{IEEEeqnarray}
%Here we use the notation of Ref. \cite{gradshteyn-ryzhik.2007}, where $F$ is defined in Eqs.\ 2.571.5, 2.571.7 (pp 179--180) and Eq.\ 8.111.2 on p. 860, and $\Pi$ is defined in Eq.\ 8.111.4, p 860.
%
%For a general discussion of elliptic functions, see also  Ref. \cite{press-etal.89}, \S 6.11, p 254.

%where this is derived using Eqs.\ 2.571.5 and 2.571.7 (pp 179--180), resp.; and $K(k) = F(\frac{\pi}{2},k)$, where $F$ is the elliptic integral of the first kind, Eq.\ 8.111.2, p 860; and $\Pi$ is the elliptic integral of the third kind, Eq.\ 8.111.4, p 860.

%apsrev4-2.bst 2019-01-14 (MD) hand-edited version of apsrev4-1.bst
%Control: key (0)
%Control: author (8) initials jnrlst
%Control: editor formatted (1) identically to author
%Control: production of article title (0) allowed
%Control: page (0) single
%Control: year (1) truncated
%Control: production of eprint (0) enabled
%

\end{document}